\documentclass[a4paper,fleqn,usenatbib]{mnras}
\usepackage{newtxtext,newtxmath}
\usepackage[T1]{fontenc}
\usepackage{ae,aecompl}

\usepackage{graphicx}	% Including figure files
\usepackage{amsmath}	% Advanced maths commands
\usepackage{verbatim}
\usepackage{units}
\usepackage{subfig}
\usepackage{pdflscape}
\usepackage{svg}
\usepackage{bm}
\usepackage{stfloats}
\usepackage{float}
\usepackage{multicol}
\usepackage{hyperref}
\usepackage{multirow}
\usepackage{wasysym}
\newcommand{\degree}{\ensuremath{^\circ}}

\title[Radio modelling of Apep]{Radio modelling of the brightest and most luminous non-thermal colliding-wind binary Apep}
\author[S. Bloot et al.]{S. Bloot,$^{1}$\thanks{E-mail: bloot@strw.leidenuniv.nl}
J.~R.~Callingham,$^{1,2}$
and B.~Marcote$^{3}$
\\
$^{1}$Leiden Observatory, Leiden University, PO\,Box 9513, 2300 RA, Leiden, The Netherlands\\
$^{2}$ASTRON, Netherlands Institute for Radio Astronomy, Oude Hoogeveensedijk 4, Dwingeloo, 7991\,PD, The Netherlands\\
$^{3}$Joint Institute for VLBI ERIC, Oude Hoogeveensedijk 4, 7991 PD Dwingeloo, The Netherlands\\
}
\date{Accepted XXX. Received YYY; in original form ZZZ}

\pubyear{2021}

\begin{document}
\label{firstpage}
\pagerange{\pageref{firstpage}--\pageref{lastpage}}
\maketitle

% Abstract of the paper
\begin{abstract}
\noindent Apep is the brightest and most luminous non-thermal colliding-wind binary by over an order of magnitude. It has been suggested from infrared observations that one of the Wolf-Rayet stars in Apep is launching an anisotropic wind. Here we present radio observations of Apep from 0.2 to 20\,GHz taken over 33 years. The spectrum reveals an extremely steep turnover in the flux density at low frequencies, where the flux density decreases by two orders of magnitude over only 325\,MHz of bandwidth. This exponential decline is best described by free-free absorption, with a turnover frequency at 0.54\,$\pm$\,0.01\,GHz. Above the turnover, the spectrum is well described by a power-law and a high-frequency cut-off likely caused by inverse-Compton cooling. The lightcurve of Apep shows significant variation over the observing period, with Apep brightening by over 50\,mJy in a span of 25 years at 1.4\,GHz. Models that assume spherical winds do not replicate all of the structure evident in the radio lightcurve. We derived a model that allows one of the winds in the system to be anisotropic. This anisotropic model recovers most of the structure of the lightcurve and is a significantly better statistical fit to the data than the spherical wind model. We suggest such a result is independent support that one of the Wolf-Rayet stars in Apep is launching an anisotropic wind. If the anisotropic wind model is correct, we predict a $\sim$25\% decrease of the 1.4\,GHz flux density of Apep over the next five years.
\end{abstract}


\section{Introduction}
\label{sec:intro}

The luminous Wolf-Rayet (WR) stars represent the final evolutionary phase in the life of the most massive stars, before undergoing a core-collapse supernova. WR stars are capable of launching high-velocity, line-driven winds that carry significant mass loss, enriching the Galactic interstellar medium \citep{Lamers1991}. The ephemeral WR phase of a star's life only lasts $\approx$\,$10^5$\,years \citep{Meynet2005}.

Massive stars, with zero-age main-sequence masses over 20\,M$_{\astrosun}$, preferentially occur in binaries with other massive stars \citep[e.g.][]{Chini2012, DeMink2013, Sana2014}. In such systems, the stellar winds launched by the two stars can collide and form a shock in the wind-collision region (WCR) of the colliding-wind binary (CWB). A recently discovered example of a CWB is Apep \citep{Callingham2019}, which is extremely bright at radio frequencies. \citet{Callingham2020} found that the two stars in Apep are both WR stars. This is rare, as the WR phase of these stars is short compared to their total lifetimes. One star is a carbon-rich WC8 type that launches a wind with a spectroscopic wind speed of 2100\,$\pm$\,200\,km\,s$^{-1}$ and a mass-loss rate of $\approx10^{-4.5}$\,M$_{\astrosun}$\,yr$^{-1}$. The second star is a nitrogen-rich WN4-6b type that launches a wind with a spectroscopic wind speed of 3500\,$\pm$\,100\,km\,s$^{-1}$ and a mass-loss rate of $\approx10^{-4.3}$\,M$_{\astrosun}$\,yr$^{-1}$ \citep{Callingham2020}. Most CWB systems consist of a WR star and an OB star, which results in a small wind momentum ratio as the WR stellar wind completely dominates over the wind from the other star. With two WR stars in the system in Apep, the momentum ratio is closer to equal, at $\eta = 0.44\pm0.08$ \citep{marcote2021}, since both stars are launching powerful winds.

Most CWB systems have a non-thermal flux density below 20\,mJy \citep{Monnier2002, debecker2013}. Generally, these low flux densities are explained by a combination of strong absorption, relatively weak shocks, and a lack of strong magnetic fields \citep{debecker2013, Monnier2002}. When \citet{Callingham2019} identified Apep, it had a flux density of 166\,$ \pm $\,15\,mJy at 1.4\,GHz, which is over an order of magnitude brighter and more luminous than all other CWB systems, making Apep the brightest and most luminous non-thermal CWB ever discovered. It is not yet clear why Apep is so radio bright. Possible explanations are that there is less absorption along our line of sight to the WCR, a stronger momentum ratio, or that the magnetic field of the WCR in Apep is significantly stronger than in other CWB systems \citep{Callingham2019}. 

The emission at radio frequencies originates from the WCR. It encodes the energetics of the stellar winds, orbital parameters of the binary, and magnetic field properties of the system \citep{Dougherty2003,Pittard2006}. The emission from the WCR is synchrotron emission from high-energy electrons. Synchrotron radiation is well described by a power-law at radio frequencies \citep{Rybicki_Lightman}. Most spectra of CWBs display a turnover at the low frequencies due to an absorption mechanism. What absorption mechanism is dominant depends on the properties of the system. There are three possible mechanisms that could cause the turnover at low frequencies. It is not yet clear which absorption mechanism is most common in CWBs, partially because most CWBs are not bright enough to study the spectrum of these sources in detail. 

One of these absorption effects is synchrotron self-absorption (SSA). SSA occurs when the high-energy electrons and the emitted synchrotron photons have a large chance of interaction \citep{Callingham2015}. The absorption cross-section for this interaction is larger at lower frequencies, which results in a low-frequency cut-off of the spectrum, with a spectral index of 2.5 or shallower. For wide CWBs, the effect SSA would have on the spectrum should be negligible \citep{Dougherty2003}.

Another possible absorption mechanism that could cause the observed turnover at low frequencies is the Tsytovich-Eidman-Razin effect, hereafter referred to as the Razin effect \citep{Hornby1966}. At low frequencies, the Razin effect suppresses the beaming effect that boosts synchrotron radiation. The suppression leads to an exponential cut-off as the frequency approaches the Razin frequency \citep{Rybicki_Lightman}, which depends on the electron density and the inverse of the magnetic field \citep{Dougherty2003}. This means that the Razin effect is relevant in systems with a high electron density or a weak magnetic field.

The spectral turnover in CWBs could also result from free-free absorption (FFA), which is caused by ionized media surrounding the synchrotron source absorbing the emitted radiation. The ionized medium is more opaque at the lower frequencies and thus causes an exponential turnover. FFA can cause an extremely steep turnover, steeper than the turnover caused by SSA or the Razin effect. The strength of this effect depends mainly on the amount of ionized medium along the line of sight. In many systems, FFA is thought to be the most likely mechanism causing the turnover \citep[e.g.][]{Pittard2_2006,Benaglia2020}. 

At the higher ($>$10 GHz) frequencies, these absorption mechanisms have very little effect. However, different cooling and aging mechanisms can cause a deviation from a power-law, such as inverse Compton (IC) cooling. IC cooling is a process where a photon scatters off a high-energy electron and gains energy from the interaction. This leads to an exponential cut-off in the spectrum of the source, which depends on the electron and radiation density \citep{Hornby1966}.

Since Apep has such a high flux density, as opposed to other radio-bright CWBs, we can conduct the most sensitive and broadband analysis of a CWB spectrum. With Apep, we have enough signal-to-noise that we can statistically distinguish between the different absorption mechanisms that could be causing a turnover at the low frequencies, as well as find a possible signature of IC cooling at the high frequencies.

Furthermore, the flux density of a CWB is not expected to be constant over time. The synchrotron flux density can vary with the separation between the two stars, as can the strength of the absorption mechanisms \citep{Dougherty2003}. Since FFA is caused by the absorption of the synchrotron emission by the ionized medium along the line of sight to the WCR, the free-free opacity is expected to change along the orbital period. \citet{williams1990} derived a model that describes the variation of the free-free optical depth along the orbit of the WCR in a CWB, assuming the winds from both of the stars are spherical. This model predicts significant flux density variation along the orbit of a CWB system, allowing us to derive orbital parameters of the system from monitoring the radio flux density. 

In Apep, \citet{Callingham2019} already identified significant variability at 843\,MHz, with the flux density nearly doubling in less than 20\,years. Tracing such variability over a longer period of time can help constrain the orbital elements of the system.
\citet{han2020} determined the orbital elements of Apep based on mid-infrared images taken by the Very Large Telescope (VLT). They found that the system has an eccentric orbit, with a period of $\sim$125\,years, and is currently close to apastron. Using radio monitoring observations of Apep, we can independently test if the orbital parameters derived from the infrared are consistent with what is evident from the radio emission of the system.

The mid-infrared data also revealed Apep has a spiral dust plume that is uniformly expanding \citep{Callingham2019}. \citet{han2020} determined, based on multiple epochs of VLT data, that the dust plume is expanding at 910\,$ \pm $\,120\,km\,s$^{-1}$. This is significantly slower than the measured spectroscopic wind speed. Such a discrepancy between the spectroscopic wind speed and the expansion of the dust plume had not previously been observed in CWB systems. \citet{Callingham2019} suggested that the difference could be caused by one of the stars launching an anisotropic wind, such as a fast polar wind and a slow equatorial wind. This is possible if the star is rotating at near-critical rotation speed, which would make Apep a long gamma-ray burst progenitor \citep[e.g.][]{Woosley1993, Thompson1994, MacFayden1999, MacFayden2001, Woosley2006, Detmers2008}. In this case, the usual models for FFA using spherical winds, such as the model by \citet{williams1990}, may not hold for this system.

To derive orbital parameters independent from the infrared, determine the dominant absorption mechanism, and to test if there is any evidence of unique wind properties in Apep, we observed Apep regularly from 2017 to 2021 using the Australia Telescope Compact Array (ATCA) and the Giant Metrewave Radio Telescope (GMRT). We also searched the ATCA archive to extend our dataset to 1994. In Section\,\ref{sec:obs}, we will discuss our ATCA and GMRT observations, and the archival data we used. Section\,\ref{sec:models} contains the modelling routine that we developed to determine the best fits to the data. In Section\,\ref{sec:results}, we analyse the radio spectrum of Apep to determine which absorption mechanism is dominant in the system. We also look at the time variability of the flux density to determine whether this is well described by spherical winds or anisotropic wind models, and if the orbital elements that best fit the data agree with the orbit determined from the mid-infrared observations.
In Section\,\ref{sec:discuss} these results are discussed. The conclusions of this work are presented in Section\,\ref{sec:concl}.

\section{Radio Observations and Data Reduction}
\label{sec:obs}

Apep has been observed by several radio telescopes, including the Australia Telescope Compact Array (ATCA), the Giant Metrewave Radio Telescope (GMRT), the Molonglo Observatory Synthesis Telescope (MOST), and the Australian Square Kilometre Array Pathfinder (ASKAP). In this section, we present the details of the observations and data reduction processes performed.
\subsection{ATCA}
From 2017 to 2021, Apep was regularly observed with the ATCA (Project ID: C3267; PI: Callingham). Apep was also serendipitously observed by other ATCA observing programs, such as those surveying the Galactic plane, providing ten archival observations from 1994 to 2016. The specifics of each ATCA observation are listed in Table\,\ref{tab:atca}. We note that observations conducted before 2010 were completed using the old 128\,MHz backend of the telescope, while observations after 2010 use the 2\,GHz Compact Array Broadband Backend \citep[CABB;][]{Wilson2011}.

For the observations conducted in L-, C-and X-band post-2016, the phase calibrator was targeted after 30 minutes on source. During the K-band observations, the phase calibrator was targeted after every 20 minutes on source, and after 50 minutes pointing calibration was performed using the primary calibrator.

Most of the observations were conducted using the array in a standard 6-km configuration, as can be seen in Table \ref{tab:atca}. However, several observations were completed in a 750\,m configuration or a 1.5\,km configuration. During the observations in May 2017, antenna CA01 was not available due to maintenance. During the observations in February 2021, there was an issue with a correlator block which led to antenna CA03 being unusable for the second observed frequency. During the observations in March 2021, antenna CA05 could not be used in L-band.

All data from the ATCA was reduced using {\ttfamily \selectfont CASA} (v\,5.7). 
In L-band, between 25\% and 55\% of the data was flagged due to radio frequency interference (RFI). In the other bands, less than 5\% was flagged. The flagging was done using the automatic flagging algorithm {\ttfamily \selectfont tfcrop}, as well as manually after inspection. Some of the short baselines had to be flagged completely in L-band owing to poor RFI conditions.

Each of the observations conducted post-2010 has a bandwidth of 2\,GHz. The band was split into smaller frequency bins to allow for more detailed spectral analysis. The L-band data was split into 18 bins with a width of 114\,MHz. The C-band was split into 7 bins, each with a width of 292\,MHz. The X-band data was split into 4 bins with a width of 512\,MHz and the K-band was split into 2 bins of 1024\,MHz per band. The width of the bins was chosen such that the spectral information in each bin is approximately the same, and to ensure each frequency bin traces roughly the same fractional bandwidth.

Gain and bandpass calibration was performed on the entire bandwidth for L-band, but per bin for the other bands. Most of the observations from our observing campaign are calibrated on PKS\,B1934-638, but the L-band data from 2017-05-11 is calibrated on PKS\,B0823-500 and all K-band data is calibrated on PKS\,B1921-293.
Phase calibration on all epochs of our observing campaign was performed using PMN\,J1534-5351. For the archival data, we used the gain and phase calibrator included in the datasets, as listed in Table\,\ref{tab:atca}.

First, gain, bandpass and leakage solutions were determined using the flux density calibrator and a time interval of 60\,s. These solutions were transferred to the phase calibrator. The combined solutions of the gain, bandpass, leakage and phase calibration were then transferred to Apep. After transferring the solutions and more flagging, the data was imaged using {\tt tclean} with a robust parameter of 0.5. We then performed a phase self-calibration step using the cleaned images as a model. This process of cleaning and self-calibrating is repeated one more time, after which the final images were made by cleaning until 5 times the theoretical noise level.

The total flux density of Apep in each image was determined using \texttt{Aegean} (v\,2.2.3) \citep{aegean_2, aegean} and BANE for the background and noise estimation. BANE creates background and noise images that can be used directly by \texttt{Aegean}. \texttt{Aegean} is a source finding program that is designed to work efficiently on large datasets. In this work, the flux density was measured using the priorized fitting function, allowing the flux density to vary but not the position of the source or the size of the beam. Many of these observations have very elongated point spread functions, meaning it was necessary to fix the location of the source as well as the shape and the size of the beam.

The uncertainties on the flux densities were determined based on the spread of amplitude values of the primary calibrator and the local rms value returned by \texttt{Aegean}, combined in quadrature.

\subsection{GMRT}

Apep was observed with the uGMRT (project ID: 35\_093; PI: Marcote) at low frequencies: band 2, centred at 200~MHz, and band 4, centred at 650~MHz.
All of our GMRT observations used the same frequency configuration of a total bandwidth of 200~MHz divided into 2048~channels, recording dual circular polarization. The specifics of each observation are listed in Table~\ref{tab:atca}. The source 3C~286 was used as amplitude calibrator and PMN\,J1603-4904 (also known as 1600$-$48 in calibrator databases) as phase-referencing calibrator source, in a phase-referencing cycle of 5~min on calibrator and 30~min on target.

% Data Reduction
All data from the uGMRT were reduced using {\tt CASA} (v 5.7) following standard procedures and by using a pipeline adapted from the CAsa Pipeline-cum-Toolkit for Upgrade Giant Metrewave Radio Telescope data REduction \citep[CAPTURE;][]{kale2021}.

The data were first flagged both automatically and manually to remove the RFI present in the band. We flagged all baselines shorter than $7~\mathrm{k\lambda}$ (i.e.\ excluding the baselines between all the core uGMRT antennas) because of a known technical issue with the GMRT Wideband Backend (GWB) system at the time of the observations. Gain and bandpass calibration was performed by using the scans on 3C~286. The absolute flux density scale was set by using the flux density predicted for 3C~286 by the standard \citet{Perley2013} model. A delay calibration was also conducted using this source with a time interval of 60~s. Phase calibration was performed on PMN\,J1603-4904, and the solutions were then transferred to Apep.
Further automatic flagging was performed and then we split the data from Apep with a channel averaging of ten. About $20\text{--}50\%$ of the data at band 4 and $\sim 75\%$ of the data at band 2 were flagged due to RFI. The final usable bandwidth at band 2 was 156~MHz, centred at 201~MHz. The significant amount of RFI in the band 2 data originally caused complete flagging of the data within the {\tt CASA} pipeline. At this step we conducted a completely manual flagging of these data in {\tt AIPS} using the {\tt TVFLG} and {\tt SPFLG} tasks to provide a completely RFI-free section of band 2.

All epochs were imaged using {\tt tclean} in CASA with a robust parameter of zero.
We then performed a self-calibration and imaging loop, using the cleaned images as a refined model for the next iteration. We conducted four phase-only self-calibration steps with decreasing integration times between 8 and 1~min, followed by one amplitude-and-phase self-calibration step with a longer integration time of 30~min. We repeated this cycle again until we obtained the final image of Apep. For band 4, this final calibrated dataset was then split into 6 bins, to obtain the maximal amount of spectral information.
We used {\tt imfit} to measure the flux density of Apep in these images, where in all cases it remained as a point-like source.

\subsection{Other radio data}
\subsubsection{Molonglo Observatory Synthesis Telescope}

Between 1988 and 2006, Apep was observed on six occasions by the Molonglo Observatory Synthesis Telescope at a frequency of 843\,MHz. These observations were mostly conducted as a part of the second epoch Molonglo Galactic plane Survey (MGPS2) \citep{MOST}. The flux densities were calculated from the MGPS2 calibrated images. This data was presented by \citet{Callingham2019}, who saw a significant variation over this time period. The measured flux density increased from 85\,$\pm$\,8\,mJy to 139\,$\pm$\,5\,mJy in a span of 18\,years. During this monitoring, the increase in flux density appeared to be linear. All flux densities reported by the MOST are available in Appendix Table\,\ref{tab:fluxes}.

\subsubsection{Rapid ASKAP Continuum survey (RACS)}

In 2020, Apep was observed with the Australia Square Kilometre Array Pathfinder (ASKAP) as a part of the Rapid ASKAP Continuum Survey (RACS) \citep{RACS_2020}. In 2020, 903 images of this survey were released. These images are all south of declination $\delta=+41\degree$ and have a bandwidth of 288\,MHz centred on 887.5\,MHz. Apep was observed as a part of this survey on 2020-04-20, with an observed flux density of 170\,$ \pm $\,20\,mJy. This flux density was calculated from the calibrated images of RACS using \texttt{Aegean}. 

\begin{table*}
\caption{\label{tab:atca} Summary of the observations used in this work. On 2019-09-11 Apep was observed in L-, C- and X-bands, and on 2019-09-12 in K-band. Since we do not expect any noticeable variability on such short timescales, we will treat this as one epoch. The primary calibrator in the table is the one used for all bands observed in that particular epoch except K-band. For K-band, PKS B1921-293 was used for all epochs. The observations above the horizontal line are those taken with the old 128\,MHz backend of the ATCA, those below the line are with the new backend.}
\begin{center}
\begin{tabular}{lrrrrrrrrr}
\hline
Epoch & Telescope &Array & Band 2 & Band 4 & L-band & C-band & X-band & K-band & Primary calibrator \cr
\hline
1994 Jan 10 & ATCA & 6A & && 4$^\prime$ & & & & PKS B0823-500 \cr
1994 Jan 22 & ATCA & 6A & && 3$^\prime$ & & & &PKS B0823-500  \cr
2004 Mar 21 & ATCA & 1.5A & &&90$^\prime$ & & &   & PKS B0823-500 \cr
2004 Mar 23& ATCA & 1.5A & && 245$^\prime$ & & &  &PKS B1934-638 \cr
2004 May 05 & ATCA & 6C & &&570$^\prime$ & & & & PKS B1934-638  \cr
\hline
2010 Dec 31 & ATCA & 6A & && & 2$^\prime$ & &  &PKS B1934-638 \cr
2011 Jan 27 & ATCA & 6A & &&29$^\prime$ & & &  & PKS B0823-500  \cr
2012 Jul 03 & ATCA & 750A & &&13$^\prime$ & & &  & PKS B1934-638 \cr
2013 Jan 27 & ATCA & 750C & &&12$^\prime$ & & &  & PKS B1934-638 \cr
2016 May 29 & ATCA & 750A & &&1$^\prime$ & & &  & PKS B1934-638 \cr
2017 May 11 & ATCA & 6A & &&180$^\prime$ & 120$^\prime$ & 120$^\prime$ & 197$^\prime$ & PKS B0823-500 \cr
2018 Oct 08 & ATCA & 6A & &&120$^\prime$ & 60$^\prime$ & 60$^\prime$ &  & PKS B1934-638\cr
2018 Nov 24 & uGMRT & --- & & $75^\prime$ & &&&& 3C~286 \cr
2018 Nov 25 & ATCA & 6B & &&120$^\prime$ & 55$^\prime$ & 55$^\prime$ &  & PKS B1934-638\cr
2019 Jan 30 & uGMRT & --- & & $70^\prime$ & &&&& 3C~286 \cr
2019 Mar 22 & uGMRT & --- & & $50^\prime$ & &&&& 3C~286 \cr
2019 Mar 23 & uGMRT & --- & $117^\prime$ && &&&& 3C~286 \cr
2019 Mar 24 & ATCA & 6A & &&120$^\prime$ & 65$^\prime$ & 65$^\prime$ &  & PKS B1934-638\cr
2019 Sep 11-12 & ATCA & 6C & &&120$^\prime$ & 67$^\prime$ & 67$^\prime$ & 108$^\prime$ & PKS B1934-638 \cr
2020 May 02 & ATCA & 6A & &&120$^\prime$ & 45$^\prime$ & 45$^\prime$ &  & PKS B1934-638\cr
2021 Feb 17 & ATCA & 750C & && & 60$^\prime$ & 60$^\prime$ & 109$^\prime$  & PKS B1934-638\cr
2021 Mar 07 & ATCA & 6D & &&88$^\prime$ & 55$^\prime$ & 55$^\prime$ &  & PKS B1934-638\cr
2021 Jun 28 & ATCA & 6B & && 122$^\prime$ & 132$^\prime$ & 132$^{\prime}$ & & PKS B1934-638\cr

\hline
\end{tabular}
\end{center}
\end{table*}

\section{Modelling Routine}
\label{sec:models}

To determine which of the different emission and absorption models best fits the radio spectrum and the lightcurve of Apep we have implemented a Bayesian inference model fitting routine. Such a routine also allows us to calculate the model parameters and uncertainties. We apply a methodology similar to that presented by \citet{Callingham2015}.

Defining the data $\boldsymbol{D}$ as a matrix containing the observed flux densities at certain frequencies and times, and the parameter vector $\boldsymbol{\theta}$ as the vector containing the values of the parameters and $M$ as the model, we can use Bayes' theorem

\begin{equation}
    \Pr(\boldsymbol{\theta}|\boldsymbol{D}, M) = \frac{\Pr(\boldsymbol{D}|\boldsymbol{\theta}, M) \Pr(\boldsymbol{\theta}| M)}{\Pr(\boldsymbol{D}|M)} ,
\end{equation}

\noindent to determine $\Pr(\boldsymbol{\theta}|\boldsymbol{D}, M)$, the posterior distribution of the model parameters. The posterior distribution represents the updated belief of the model parameters given the data. $\Pr(\boldsymbol{D}|\boldsymbol{\theta}, M) \equiv \mathcal{L}(\boldsymbol{\theta})$ is the likelihood function. This is the likelihood of observing this set of data given the model and parameters. $\Pr(\boldsymbol{\theta}| M) \equiv \Pi(\boldsymbol{\theta})$ is the prior information for a given model and $\Pr(\boldsymbol{D}|M) \equiv Z$ is the Bayesian evidence, an indication of how well the model predicts the observed data.
\begin{table*}

\caption{\label{tab:results} Results of the spectral fitting to the data taken by the ATCA on 2019-03-24 and 2019-09-12 combined with the data from the uGMRT observations on 2019-03-22 and 2019-03-23. $S_{\mathrm{norm}}$ is a normalisation constant for the flux density, $\alpha$ is the spectral index of the synchrotron emission, and $\beta$ is defined according to $\alpha = -(\beta-1)/2$. $\nu_{\mathrm{peak,FFA}}$, $\nu_{\mathrm{peak,SSA}}$ and $\nu_{\mathrm{Razin}}$ are the turnover frequencies caused by FFA, SSA, and the Razin effect, respectively. $\nu_{\mathrm{break,IC}}$ is the break frequency caused by IC cooling. ln(Z) is the Bayesian evidence value of the fit.} 
\begin{center}

\def\arraystretch{1.5}
\resizebox{2.1\columnwidth}{!}{
\begin{tabular}{lcccccccc}

\hline
Model & $S_{\mathrm{norm}}$ (mJy) & $\alpha$ & $\beta$ & $\nu_{\mathrm{peak,FFA}}$ (GHz) & $\nu_{\mathrm{peak,SSA}}$ (GHz) & $\nu_{\mathrm{Razin}}$ (GHz) & $\nu_{\mathrm{break,IC}}$ (GHz) &ln(Z) \cr
\hline

FFA  & 353  $\pm$ 4 & -0.73  $\pm$ 0.01 &  & 0.66  $\pm$ 0.02 &  &  & & 94.52 $\pm$ 0.02 \cr
SSA & 246  $\pm$ 2 & -0.71  $\pm$ 0.01 & 2.42  $\pm$ 0.01 &  & 1.01  $\pm$ 0.02 &  & & 3.94 $\pm$ 0.02 \cr
Razin & 405  $\pm$ 2 & -0.99  $\pm$ 0.01 &  &  &  & 1.39  $\pm$ 0.02 & & 80.33 $\pm$ 0.02 \cr
FFA + IC & 314  $\pm$ 4 & -0.52  $\pm$ 0.01 &  & 0.54 $\pm$ 0.01 &  &  & 36  $\pm$ 2& 186.72 $\pm$ 0.02 \cr
SSA + IC & 236  $\pm$ 2 & -0.52  $\pm$ 0.02 & 2.03  $\pm$ 0.02 &  & 0.89  $\pm$ 0.02 &  & 36 $\pm$ 2& 103.51 $\pm$ 0.02 \cr
Razin + IC & 405  $\pm$ 2 & -0.99  $\pm$ 0.01 &  &  &   & 1.39  $\pm$ 0.02 & (4.7 $\pm$ 0.3)$\times 10^4$ & 80.26 $\pm$ 0.09 \cr
\hline
\end{tabular}
}
\end{center}
\end{table*}
We can assume that the data collected from the different instruments used in this study are independent. Although most of the data is from the ATCA, we can again assume that the data points in the same band are independent since the receivers have a response that ensures that the different channels are independent. We can also assume that the uncertainties on the measurements $(\sigma_n)$ are normally distributed. This means that we can define the joint log likelihood function as 

\begin{equation}
    \ln \mathcal{L}(\boldsymbol{\theta}) = - \frac{1}{2} \sum_n \left[ \frac{(D_n - S_{\nu}(\nu_n))^2}{\sigma_n^2} + \ln(2\pi \sigma_n^2) \right],
\end{equation}

\noindent where $D_n$ is the observed flux density at frequency $\nu_n$, and $S_{\nu}(\nu_n)$ is the flux density predicted by the model $M$ with parameters $\boldsymbol{\theta}$ at $\nu_n$. This log likelihood function is then convolved with uniform priors for each parameter, based on physical constraints. For example, flux density can not be negative.

The most efficient way to maximize the likelihood of a model is to use Markov chain Monte Carlo (MCMC) methods. We implemented an affine-invariant ensemble sampler \citep{Goodman2010} using the Python package {\tt emcee} (v\,3.0.2) \citep{ForemanMackey2013}. This means that the algorithm uses an ensemble of ``walkers'' that is relatively insensitive to covariance between the parameters. We used the simplex algorithm \citep{Nelder1965} to direct the walkers to the maximum likelihood.

In Bayes' theorem, Bayesian evidence is required to normalize the posterior over the volume of the prior. This is defined by

\begin{equation}
    Z = \int \int ... \int \mathcal{L}(\boldsymbol{\theta}) \Pi(\boldsymbol{\theta}) d\boldsymbol{\theta},
\end{equation}

\noindent where the dimensionality of the integration is equal to the number of free parameters in the model. To calculate this value, we use the algorithm \textsc{MultiNest} (v\,3.10) \citep{Feroz2013}, which implements nested sampling. The algorithm initializes a number of points uniformly sampling the prior space, and contracts the distribution around points of high likelihood by discarding the points with the lowest likelihood and re-initialising them randomly in the smaller prior space. This is repeated until the region of maximum likelihood is found.

To find the optimal set of parameters for a given model and the data, we first initialise 100 {\tt emcee} walkers and let them walk for 1000 steps. We analyse the last 500 steps to find the best values for the parameters and their uncertainties, accounting for a burn-in phase. From these parameter estimations, we calculate a new prior range for \textsc{MultiNest}. The lower and upper limits of this prior range are set to 10\% below and above the best parameter value from the MCMC results.

To assess which model best describes the data, we consider the difference in evidence values calculated by \textsc{MultiNest}. The best-fitting model between two competing models that are \textit{a priori} equally likely to describe the data is evaluated by considering the ratio of their evidence values. Expressed in log space, we write this as $\Delta \ln(Z) = \ln(Z_2) - \ln(Z_1)$. Using an updated version of the Jefferys scale \citep[e.g.][]{Kass1995, Scaife2012, Callingham2015}, $\Delta \ln(Z) \ge 3$ is strong evidence that $M_2$ is significantly better at describing the data than $M_1$. $1<\Delta \ln(Z) < 3$ is moderate evidence that $M_2$ describes the data better than $M_2$, and $0\le\Delta \ln(Z) \le 1$ is inconclusive. 

\section{Results}
\label{sec:results}

\subsection{Spectral modelling}
\label{sec:spectral}

Apep is extremely bright at radio frequencies when compared to other CWBs, which allows us to conduct a sensitive spectral analysis. As described in Section \ref{sec:obs}, each band of the ATCA observations is split into bins with each roughly the same amount of spectral information to give us the maximal amount of data points.

To study the full spectral energy distribution (SED) of Apep, we combine the ATCA observations at L-, C- and X-band taken on 2019-03-24 with the ATCA K-band observation from 2019-09-12. While these observations are taken roughly six months apart, combining them does not bias the analysis as there is no significant variation in the K-band flux density across four years of our monitoring observations. Combining the ATCA data with the flux densities from the band 2 and 4 uGMRT data taken only days earlier (2019-03-22 and 2019-03-23) results in the SED shown in Figure \ref{fig:sed}. The data shows an extremely steep turnover at the low frequencies, where the spectrum drops from $143 \pm 14$\,mJy at 583\,MHz to $3.1 \pm 0.8$\,mJy at 255 MHz. The turnover can be potentially modelled by several different absorption processes, as outlined in Section \ref{sec:intro}. The absorption mechanisms considered here are free-free absorption, synchrotron self-absorption, and the Razin effect. The intrinsic synchrotron emission of the system can be described by a standard non-thermal power-law spectrum. 

We can also safely neglect the contribution of thermal flux density in the modelling of the radio spectrum because the thermal emission from the stars is far below the measured flux density of Apep at all of our observing frequencies. For example, the expected thermal flux density can be calculated using 
\begin{equation}
    S_{\nu, \mathrm{thermal}} = 23.2 \left(\frac{\dot{M}}{\mu \mathrm{v}_{\infty}}\right)^{4/3} \frac{\nu^{2/3}}{D^2} \gamma^{2/3} g^{2/3} Z^{4/3} \, \mathrm{ Jy},
\end{equation}
where $S_{\nu, \mathrm{thermal}}$ is the thermal flux density of a single star, assuming a spherical wind structure. Any deviation from this structure should decrease the flux density. $\dot{M}$ is the mass-loss rate in M$_{\astrosun}$\,yr$^{-1}$, $\mu$ is the mean atomic weight, $\mathrm{v}_{\infty}$ is the terminal wind velocity in km\,s$^{-1}$, $\nu$ is in Hz, $D$ is the distance to the star, $\gamma$ is the number of free electrons per nucleon, $g$ is the Gaunt factor, and $Z$ is the mean ionic charge \citep{Wright1975}. \citet{Callingham2020} found a mass-loss rate of $10^{-4.5}$\,M$_{\astrosun}$\,yr$^{-1}$ and a terminal wind velocity of 3500\,km\,s$^{-1}$ for the WN star in Apep, and a distance of $\approx$2\,kpc to the system. Assuming a wind composition typical of WN stars, $\gamma=1$, $g=1$ and $Z=1$ \citep{Leitherer1997}, we find that across the observed frequency range, the thermal flux is always less than 1\,mJy, far below the $\approx$30\,mJy measured for the system at our highest frequencies. Furthermore, VLBI observations at 2\,GHz, with a rms noise of 40\,$\mu$Jy, did not detect any emission from the stars themselves \citep{marcote2021}.

\subsubsection{Free-free absorption}
\begin{figure}
    \centering
    \includegraphics[width=\columnwidth]{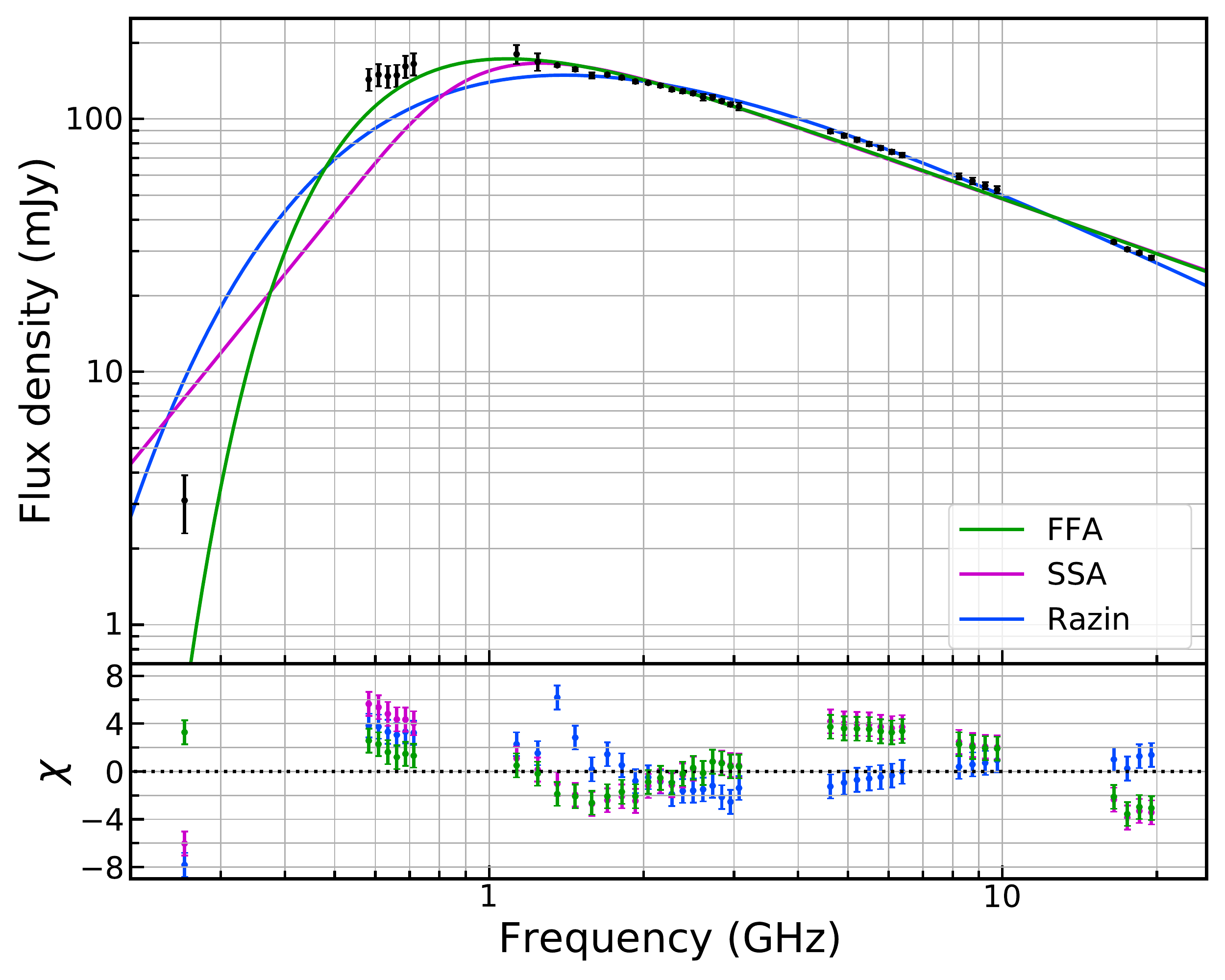}
    \caption{The top plot shows a SED of the ATCA data from March and September 2019 combined with the GMRT data from March 2019 in black. The green line shows the best fit of the FFA model to the data. The purple line shows the best fit of the SSA model and the blue line shows the best fit of the Razin effect model. The bottom plot shows the residuals of each fit plotted in the same colour. The vertical axis shows the $\chi$-value, defined as the difference between the data and the model divided by the uncertainty on the measurement.}
    \label{fig:sed}
\end{figure}

Free-free absorption (FFA) is the absorption of radiation by an external ionized screen relative to the emitting electrons. 
In this case, the optical depth $\tau_{\mathrm{FFA}}$ due to FFA can be described by \citep{Mezger1967}
\begin{equation}
\label{eq:ffa}
    \tau_{\mathrm{FFA}} \approx 8.24 \times 10^{-2} \nu^{-2.1} T_e^{-1.35} \int n_e^2 \mathrm{d}l,
\end{equation}
where $\nu$ is the frequency in GHz, $T_e$ is the electron temperature in K, $n_e$ is the electron density in cm$^{-3}$ and $l$ is the distance through the source in pc.
We can define a characteristic frequency $\nu_{\mathrm{peak,FFA}}$ such that
\begin{equation}
    \tau_{\mathrm{FFA}} = \left(\frac{\nu}{\nu_{\mathrm{peak,FFA}}}\right)^{-2.1},
\end{equation}
where $\nu_{\mathrm{peak,FFA}}$ is the frequency at which the optical depth is unity.

The resulting spectrum, including the synchrotron emission power-law, is described by
\begin{equation}
\label{eq:ffa_spectral}
    S_{\nu} = S_{\mathrm{norm}} \left(\frac{\nu}{\nu_{\mathrm{peak, FFA}}}\right)^{\alpha} e^{-\tau_{\mathrm{FFA}}},
\end{equation}
where $S_{\mathrm{norm}}$ characterises the intrinsic synchrotron flux density.

This model predicts an exponential turnover at lower frequencies. At frequencies higher than $\nu_{\mathrm{peak,FFA}}$, the spectrum approaches a standard power-law.

The best fit of this model to the radio spectrum of Apep is shown in Figure\,\ref{fig:sed}. The log of the Bayesian evidence value for this model and the best-fitting parameters can be found in Table\,\ref{tab:results}. This model provides a good fit to the data in the middle of the considered frequency range, but at the higher frequencies it no longer describes the data well. At frequencies below 1\,GHz, the model underestimates the observed flux density of Apep.

\begin{figure}
    \centering
    \includegraphics[width=\columnwidth]{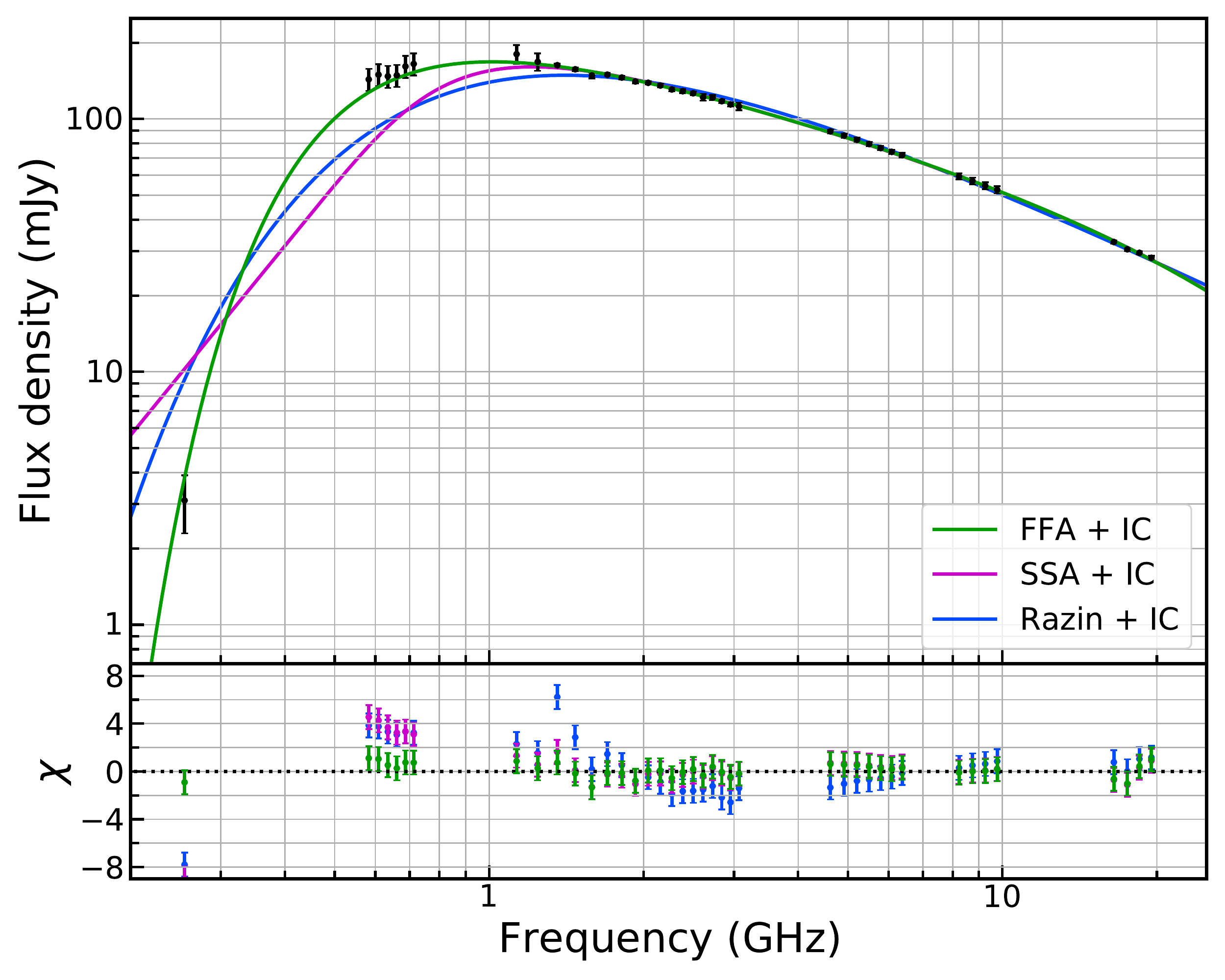}
    \caption{This figure is identical to Figure\,\ref{fig:sed}, except that here the combinations with IC cooling are plotted. The ATCA and uGMRT data is plotted in black. The green line shows the best fit for FFA combined with IC cooling. The purple line is the best fit for SSA combined with IC cooling and the blue line is the best fit for the Razin effect combined with IC cooling.}
    \label{fig:sed_ic}
\end{figure}
\subsubsection{Synchrotron self-absorption}

Synchrotron self-absorption (SSA) is the process where a high-energy electron emits a photon through synchrotron radiation, and later in its orbit reabsorbs this photon. The cross-section of this interaction is larger at lower frequencies. The optical depth of this interaction can be described by
\begin{equation}
    \tau_{\mathrm{SSA}} = \left(\frac{\nu}{\nu_{\mathrm{peak,SSA}}}\right)^{-(\beta+4)/2},
\end{equation}
where $\nu_{\mathrm{peak,SSA}}$ is the frequency at which the source becomes optically thick and $\alpha = -(\beta-1)/2$ \citep{Tingay2003}.
The full spectrum produced by synchrotron emission with SSA included is modelled by
\begin{equation}
    S_{\nu} = S_{\mathrm{norm}} \left( \frac{\nu}{\nu_{\mathrm{peak,SSA}}}\right)^{-(\beta -1)/2} \left(\frac{1-e^{-\tau_{\mathrm{SSA}}}}{\tau_{\mathrm{SSA}}}\right).
\end{equation}

The turnover caused by this effect is generally less steep than the turnover caused by FFA, limited to a maximum slope of 2.5 \citep{Kellermann1969}. The parameters for this model fit are provided in Table\,\ref{tab:results}. Figure \ref{fig:sed} shows that although the SSA model is a good fit for most of the optically-thin part of the spectrum, it does not describe the data well at the high- or low-frequency end. As can be seen from the figure and from the difference in evidence values, the SSA model is a significantly worse fit to the data than the FFA model.

\subsubsection{Razin effect}

The Razin effect is a plasma effect that suppresses the beaming effect that is needed for synchrotron emission below the Razin frequency $\nu_{\mathrm{Razin}} = 20 \frac{n_e}{B}$ \citep{Hornby1966}. 
The spectrum caused by this effect can be described by
\begin{equation}
    S_{\nu} = S_{\mathrm{norm}}\nu^{\alpha} e^{-\nu_\mathrm{{Razin}}/\nu}.
\end{equation}
The best fit of this model to the data is shown by the blue line in Figure\,\ref{fig:sed}. The parameters are listed in Table\,\ref{tab:results}. This model describes the data better than SSA, but not as well as FFA. It is the only model that describes the high-frequency data somewhat accurately, but it does not agree with the data at the low frequencies.

\subsubsection{Inverse-Compton cooling}

None of the models so far have described the entire spectrum well, particularly at frequencies below 1\,GHz and above 10\,GHz. One possible explanation for this is a high-frequency cut-off caused by inverse-Compton (IC) cooling, which is not accounted for in the standard models. IC scattering is a process where a photon scatters off a moving electron, causing its energy to increase. The photon is scattered to a higher frequency and the electron loses energy. IC cooling often occurs in environments with a strong radiation field \citep{Kellermann1969}. Since Apep is a double WR binary, it most likely has a strong radiation field and it is therefore important to consider the effect of IC cooling on the spectrum.

According to \citet{Komissarov1994}, IC cooling leads to a high-frequency exponential cut-off that can be modelled with a break frequency $\nu_{\mathrm{break, IC}}$. A spectrum with only IC cooling and no absorption effects is given by
\begin{equation}
     S_{\nu} = S_{\mathrm{norm}}\nu^{\alpha} e^{-\nu/\nu_{\mathrm{{break, IC}}}}.
\end{equation}
We combine IC cooling with FFA, SSA, and the Razin effect to see if the deviation of the models from the data at the high frequencies can be explained by this mechanism. The best-fitting parameters for each of these models can be found in Table\,\ref{tab:results} as well as their evidence values. Figure\,\ref{fig:sed_ic} shows the models that also include IC cooling. From the evidence values and Figure\,\ref{fig:sed_ic} it is clear that the addition of IC cooling to the FFA and SSA models substantially improves the fit. The addition of IC cooling to the Razin effect does not improve the fit as the break frequency appears unconstrained. 

Based on the evidence values, the best-fitting model is FFA combined with IC cooling. We discuss the implications of FFA providing the best fit to the radio spectrum of Apep in Section\,\ref{sec:discuss}.

Note that we also fit every possible combination of the absorption models to the data, such as FFA+SSA and FFA+Razin. The majority of these combinations did not accurately fit the data or were significantly worse fits than FFA combined with IC cooling. As such, we exclude considering combination absorption models further.

\subsection{Temporal modelling}

The results of the spectral modelling are derived from one epoch of observation. However, we have observations of Apep that span 27 years with the ATCA. Combining this with the observations taken by the Molonglo Observatory Synthesis Telescope \citep{Callingham2019} and ASKAP \citep{RACS_2020} provides us with a comprehensive picture of how Apep's flux density has varied over the last 33 years.

We apply an FFA model to describe the observed time variability of the flux density because we know from our spectral modelling that the spectrum of Apep is well described by a combination of FFA and IC cooling. Note we exclude the impact of IC cooling in our temporal modelling since IC cooling only impacts the spectrum of Apep at frequencies higher than 5\,GHz. As we are only looking at the variation between 800\,MHz and 3.2\,GHz, we can safely neglect IC cooling in the temporal modelling. Note that limiting the modelling to data below 3.2\,GHz will cause some discrepancy between the spectral index derived by the spectral and temporal modelling. 

\subsubsection{Spherical winds model}
\begin{table}

\caption{\label{tab:orbit} The model parameters derived from fitting both the spherical winds and anisotropic wind models to the radio lightcurve of Apep. $\xi$ is a constant proportional to $n_e^2(a)$, $s$ is the power-law index of the intrinsic synchrotron variation, $c$ is the constant in Equation\,\ref{eq:gaussian_c}, and $\omega_{\mathrm{wind}}$ is an angle that represents the orientation of the slow wind.} 
\begin{center}
\def\arraystretch{1.2}

\begin{tabular}{lll}
\hline
Model & Spherical winds & Anisotropic wind \cr
\hline
$S_{\mathrm{norm}}$ & 200 $\pm$ 2 mJy & 191 $\pm$ 2 mJy \cr 
$\alpha$ & -0.65 $\pm$ 0.01 & -0.66 $\pm$ 0.05 \cr
Period  & 142 $\pm$ 9 yr & 116 $\pm$ 10 yr\cr
Year of periastron passage & 1966 $\pm$ 6 & 1926 $\pm$ 5 \cr
Eccentricity ($e$) &  0.71 $\pm$ 0.08 & 0.68 $\pm$ 0.09 \cr
Inclination ($i$) & $\pm$ 21 $\pm$ 6\degree & -23 $\pm$ 6\degree \cr 
Argument of periastron ($\omega$) & -8 $\pm$ 7 \degree & 16 $\pm$ 10\degree\cr
$\xi$ & 0.26 $\pm$ 0.04 & 14 $\pm$ 6 \cr
$s$ & -0.01 $\pm$ 0.01 & -0.17 $\pm$ 0.07 \cr
$c$  & - & 1.5 $\pm$ 0.7 \cr
Inclination of the wind ($i_{\mathrm{wind}}$)& -& 22 $\pm$ 6\degree \cr
$\omega_{\mathrm{wind}}$&-& -23 $\pm$ 10 \degree\cr
ln(Z) & 621.8 $\pm$ 0.1 & 644.6 $\pm$ 0.1\cr
\hline
\end{tabular}

\end{center}
\end{table}

Assuming that the stellar winds from the two WR stars are both spherical, the variation in the free-free optical depth with time can be described by the model derived by \citet{williams1990}.

Assuming a spherical wind with ${n_{e} \propto d^{-2}}$ \citep[e.g.][]{Waters1988}, where $d$ is the distance from the source of the wind divided by the semi-major axis, the free-free optical depth is given by Equation\,\ref{eq:ffa}.
The optical depth can be written as 
\begin{equation}
\label{tau_step}
\tau_\mathrm{FFA} = C_{\mathrm{FFA}} a^{-1} \int d^{-4} \mathrm{d}s,
\end{equation}

\noindent where $C_{\mathrm{FFA}}$ is defined as $C_{\mathrm{FFA}} = \xi \left(\frac{\nu}{\nu_{\mathrm{ref}}}\right)^{ -2.1}$ \citep{Kennedy2010}. We use $\nu_{\mathrm{ref}}$=1.38\,GHz in this work. $\xi$ is a constant proportional to $n_e^2(a)$. The integral of Equation\,\ref{tau_step} is evaluated in Appendix\,\ref{app:williams}. The solution of this integral combined with Equation\,\ref{tau_step} gives us an expression for how the FFA optical depth varies in terms of the orbital parameters of the system. Namely,

\begin{equation}
\label{eq:time_ffa}
    \begin{split}
         \tau_{\mathrm{FFA}}(f, \nu) = C_{\mathrm{FFA}}&\frac{\mathrm{sec}(i)}{2 \Delta r^3 \mathrm{cos}^3(\omega + f)}\bigg[\mathrm{sin}(\omega+f)\mathrm{cos}(\omega+f)\mathrm{tan}(i)+\\
         & \frac{1+\mathrm{tan}^2(i)}{\sqrt{\Delta}}\mathrm{arctan}\left(\frac{-\sqrt{\Delta}}{\mathrm{tan}(\omega+f)\mathrm{tan}(i)}\right)\bigg],
    \end{split}
\end{equation}

\noindent where $i$ is the inclination, $r$ is the separation between the dominant WR star and the WCR normalised by the semi-major axis, $\omega$ is the argument of periastron, $f$ is the true anomaly, and $\Delta = 1+\mathrm{tan}^2(\omega+f)+\mathrm{tan}^2 (i) $.

\begin{figure}
    \centering
    \includegraphics[width=\columnwidth]{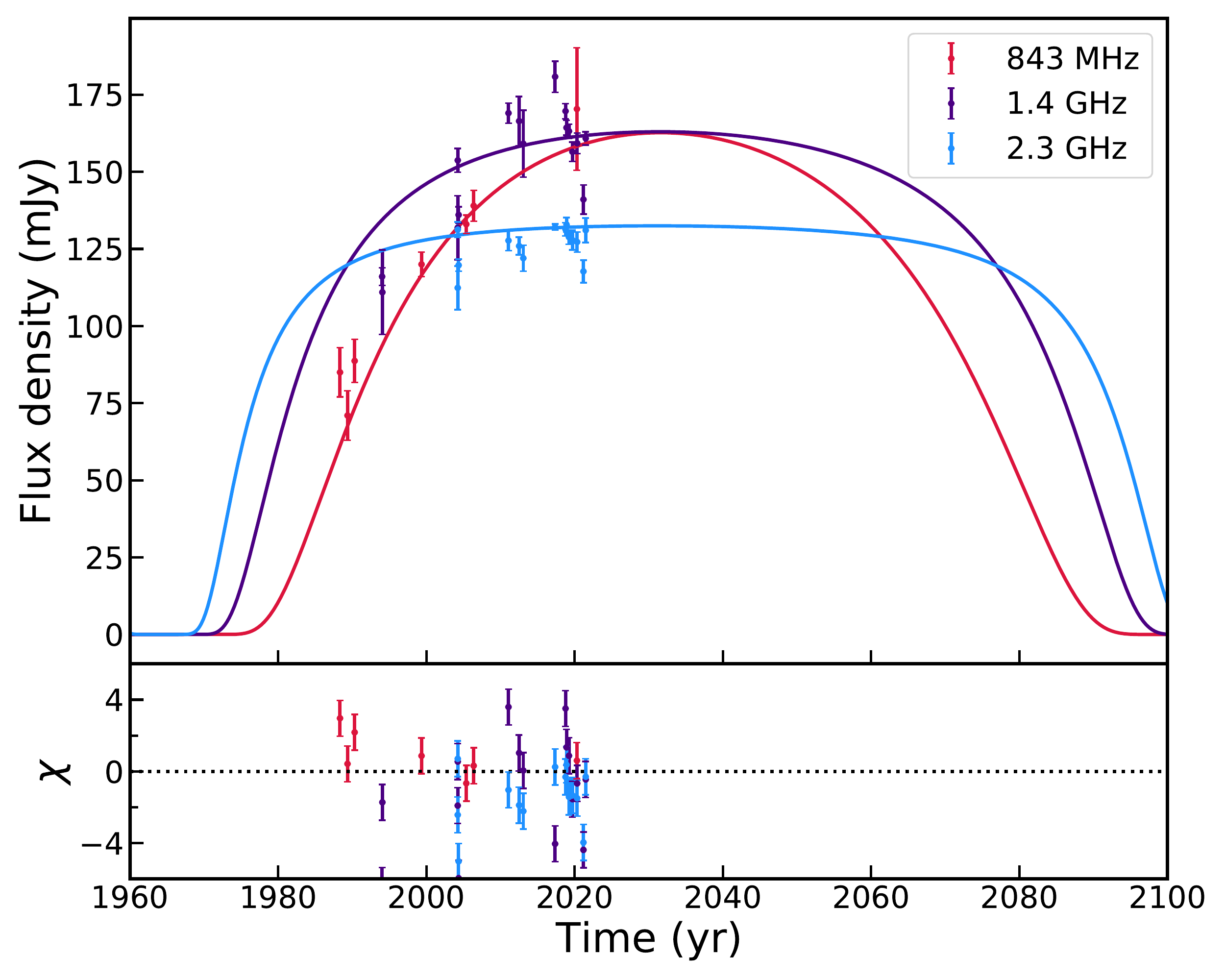}
    \caption{The best fit of the spherical wind model to the data with the prior range of the fit set to the orbital elements and associated uncertainties found in the infrared by \citet{han2020}. The top plot shows the variation of the flux density of three frequencies over time. The red line and points describe the modelled lightcurve and the data at 843\,MHz. The dark purple and blue lines represent the model and data at 1.4\,GHz and at 2.3\,GHz, respectively. The bottom plot shows $\chi$, the residuals of the fit divided by the uncertainty on the measurements. Note that the most recent red data point is from RACS at 887\,MHz.}
    \label{fig:no_disk_han}
\end{figure}
\begin{figure}
    \centering
    \includegraphics[width=\columnwidth]{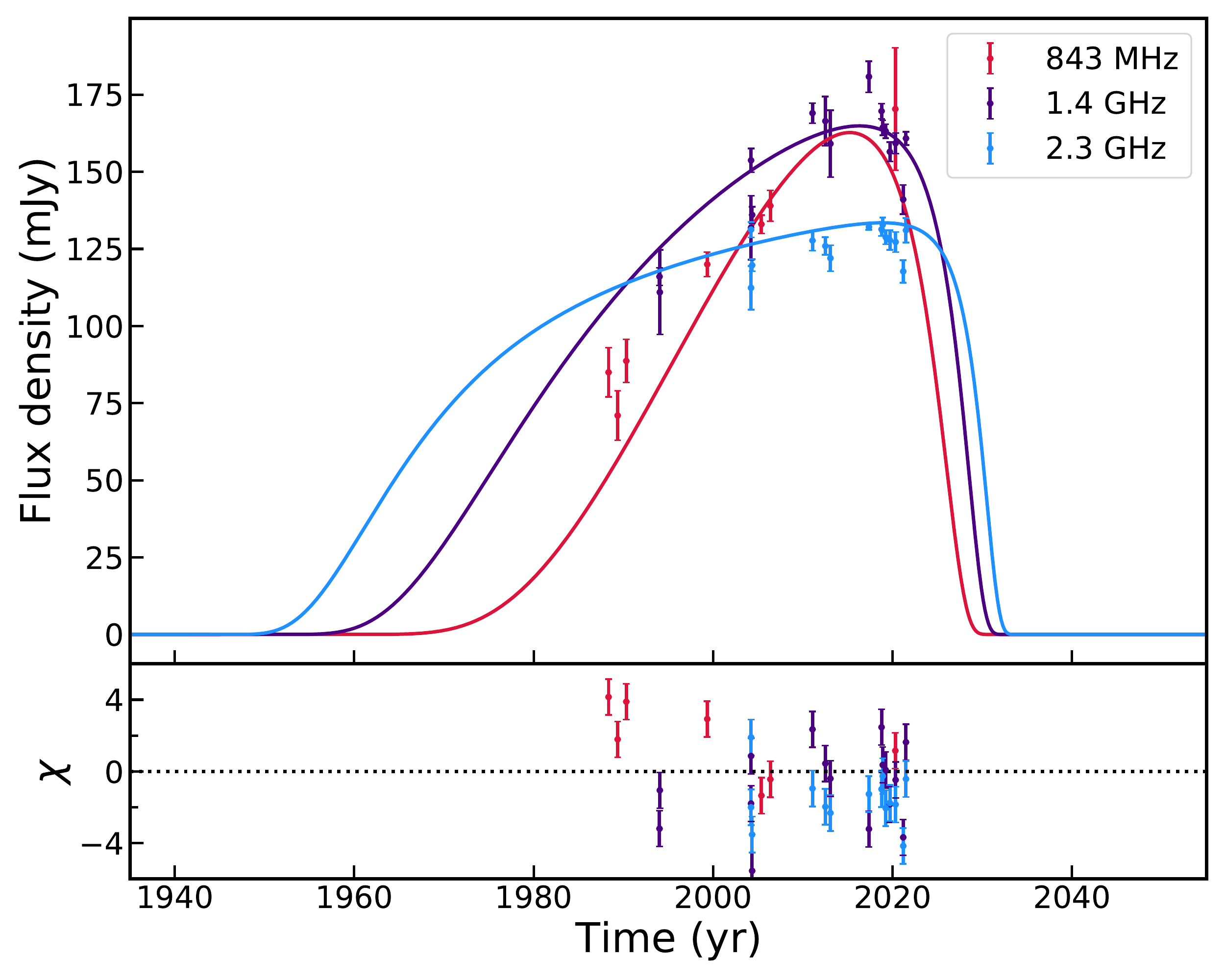}
    \caption{The best fit of the anisotropic wind model to the data with the prior range based on the orbital elements and associated uncertainties found by \citet{han2020}. The top plot shows the flux density over time. The red line and points describe the modelled lightcurve and the data, both at 843\,MHz. The dark purple line shows the model and the data at 1.4\,GHz and the blue line shows the model and the data at 2.3\,GHz. The bottom plot shows $\chi$, the residuals of the fit divided by the uncertainty on the measurements. Note that the most recent red data point is from RACS at 887\,MHz.}
    \label{fig:disk_han}
\end{figure}
Along with this expression of the orbital variation of optical depth, we need to combine it with how the orbital variation of intrinsic synchrotron flux density to accurately model the lightcurve of Apep. \citet{Dougherty2003} found that the intrinsic synchrotron emission should vary over time as 
\begin{equation}
\label{eq:syn_vary}
    S_{\nu}(d) = S_{\nu}(a) d^{-s},
\end{equation}
where ${S_{\nu}(a)}$ is the intrinsic synchrotron emission when the separation between the WCR and the dominant WR star is equal to ${a}$, and ${s}$ is a power-law index. \citet{Dougherty2003} suggest $s$ should be $-0.5$ to describe the non-thermal flux density of the WCR for most CWB systems.

Combining Equations\,\ref{eq:ffa_spectral}, \ref{eq:time_ffa}, and \ref{eq:syn_vary}, how the flux density of Apep varies as a function of time and frequency is completely given by
\begin{equation}
    S(\nu, f) = S_{\mathrm{norm}} \left(\frac{\nu}{\nu_{\mathrm{ref}}}\right)^{\alpha} e^{-\tau_{\mathrm{FFA}}(f, \nu)}d^{-s}.
\end{equation}

We fit this model to the data between 800\,MHz and 3.2\,GHz from all epochs. We exclude the ATCA observations from May 2016 due to poor image quality caused by the extremely short observation time. 

First, as a test of the orbital parameters derived from the infrared data on the system, we set the prior limits to the orbital elements and associated uncertainties derived by \citet{han2020}. The best fit with these priors is shown in Figure \ref{fig:no_disk_han}. The parameters and uncertainties are shown in Table\,\ref{tab:orbit}. The log of the evidence value for this fit is $621.8 \pm 0.1$. Although this model reproduces some of the structure, it does not describe all of the observed features of the lightcurve, as can be seen from the residuals. In particular, it does not trace the shape of the data at 1.4\,GHz, especially the decrease in flux density we observe in the last five years.

To test whether the spherical wind model fit could be improved without being limited by the \citet{han2020} priors, we fit the same model again with wide prior ranges, only limited by the physical limits of the orbital parameters.
The best fit of this model using the wider prior range has a higher ln(Z) value than the model with the prior range based on \citet{han2020} and replicates most of the observed variation. However, the fit requires a period of $\sim$40\,years and an eccentricity of $\sim$0.05. This is in direct conflict with the measurements in mid-infrared by \citet{han2020}, which show that the morphology of the dust spiral requires a high eccentricity of $\sim$0.7 and an orbital period of $\sim$100\,yr. An orbit with an eccentricity of 0.05 is therefore only possible if there is a third object in the system causing the wind collision. However, very long baseline interferometry (VLBI) imaging by \citet{marcote2021} shows that the WCR is best described as occurring between two WR stars and in the location we would expect based on the momentum ratio of those two stars. The position of the two WR stars derived from the VLBI position is consistent with that derived from the infrared. Furthermore, there is no evidence of a third companion in the optical or near-infrared spectrum \citep{Callingham2020}. We consider the presence of a third object causing the WCR implausible. Therefore, we conclude that the standard spherical wind model from \citet{williams1990} does not reproduce all of the structure in Apep's radio lightcurve, especially if we want the orbital parameters to be consistent with what has been derived from infrared and VLBI observations.

\subsubsection{Anisotropic wind model}

The failure of the standard spherical wind model to replicate all of the observed features in the radio lightcurve of Apep suggests we need to develop a more sophisticated model. As mentioned in Section\,\ref{sec:intro}, there is evidence from the infrared that one of the WR stars may be launching an anisotropic wind. Therefore, we attempt to build this feature into the model for the radio lightcurve. 

To model the anisotropic wind from one of the WR stars in Apep, we start from Equation\,\ref{eq:ffa} and rewrite it using $d^2 = R^2+z^2$, where $R$ is the component of $d$ that lays in the plane of the sky and $z$ is the component of $d$ that lies orthogonal to $R$, both in units of the semi-major axis. We can now integrate over $z$ along the line of sight to the source, as shown in the Appendix Figure\,\ref{fig:orbit_anisotropic}.

In a system with an anisotropic wind, the wind density cannot be described with $n_e \propto d^{-2}$ as before. Instead, we model the slow equatorial wind with a vertical Gaussian distribution \citep[e.g.][]{Carciofi2006, Bjorkman1997}:
\begin{equation}
\label{eq:gaussian}
    n_{e}(R^{\prime},z^{\prime}) = n_e(a,0) \left({R^{\prime}}\right)^{-3.5} \exp\left(-\frac{z^{\prime 2}}{2H^2(R^{\prime})}\right),
\end{equation}
where $R^{\prime}$ is the component of the position vector that lies in the plane of the slow wind, in units of the semi-major axis. $z^{\prime}$ is the component of the position vector orthogonal to $R^{\prime}$, also in units of the semi-major axis. $H(R^{\prime})$ is the scale height of the wind, defined by
\begin{equation*}
    H(R^{\prime}) = \frac{\mathrm{v}_s}{\mathrm{v}_c}{R^{\prime 3/2}},
\end{equation*}
where $\mathrm{v}_s$ is the isothermal sound speed and $\mathrm{v}_c$ is the critical speed of the star. We can write this as
\begin{equation}
\label{eq:gaussian_c}
    n_{e}(R^{\prime},z^{\prime}) = n_e(a,0) \left({R^{\prime}}\right)^{-3.5} \exp\left(-c \frac{z^{\prime 2}}{R^{\prime 3}}\right),
\end{equation}
where $c=\frac{1}{2}\left(\frac{\mathrm{v}_c}{\mathrm{v}_s}\right)^2$.

We combine Equation\,\ref{eq:ffa} and \ref{eq:gaussian_c} to calculate the free-free optical depth. We set the lower limit of the integral to the $z$-coordinate of the WCR and the upper limit to infinity. The full derivation of this model is included in Appendix\,\ref{app:disk}.

We fit this model to the data, again using the prior range based on the orbital elements by \citet{han2020}. We find that this model describes the data at 1.4\,GHz better than the spherical model, as can be seen in Figure\,\ref{fig:disk_han}. The parameters of the model with an anisotropic wind are shown in Table\,\ref{tab:orbit}. The orbital elements are in agreement with those found by \citet{han2020}, at least within the quoted uncertainties. The Bayesian evidence value of this model is significantly higher than that of the model with spherical winds. The fit to the data at 843\,MHz and 2.3\,GHz is not a significant improvement over the spherical winds model. However, it captures the decrease in flux density at 1.4\,GHz observed since 2017. We note that we also get an identical fit if we do not set the prior range to that derived by \citet{han2020}. 

\section{Discussion}
\label{sec:discuss}

\subsection{Spectral analysis}

We find that the model that best describes the 0.2 to 20\,GHz spectrum of Apep is synchrotron emission with a spectral index of $-0.52\pm0.02$, combined with FFA and IC cooling. If IC cooling is not included in the model, the spectral index is overestimated, as can be seen in Table\,\ref{tab:results}.
A spectral index of $\sim -0.5$ indicates that the power-law index of the energy distribution of the electrons is $p\approx2$.
This agrees with test particle theory, which predicts $p=2$ for a strong shock with a compression ratio of 4 \citep[e.g.][]{Axford1977, Bell1978, Blandford1978}.
Most non-thermal CWBs have a spectral index between $-0.3$ and $-0.8$ \citep[and references therein]{debecker2013}, so a spectral index of $-0.52$ is not unusual. We also find the FFA turnover frequency is 0.54\,$\pm$\,0.01\,GHz. This is similar to the FFA turnover frequency in the periodic dust-producer WR\,140, which \citet{Benaglia2020} found to be around 0.6\,GHz.

We can statistically conclude that the low-frequency turnover in the spectrum of Apep is not caused by SSA or the Razin effect, as shown in Section\,\ref{sec:spectral}. With FFA causing the turnover in the radio spectrum of Apep, we can put limits on the emission measure and electron density of the ionised, absorbing media.

We assume that the temperature of the ionised electrons is at least as hot as the photosphere of a WR star $T_e = 10^4$\,K \citep[e.g.][]{Crowther2007}. This is standard practice in assessing the emission measure of supernova remnants \citep{Callingham2016}. An estimate of the emission measure (EM) is therefore given by:
\begin{equation}
    \mathrm{EM} = \int n_e^2 \mathrm{d}l = \frac{ \nu_{\mathrm{peak,FFA}}^{2.1} T_e^{1.35}}{8.24 \times 10^{-2}} \approx (8.4 \pm 0.2) \times 10^5 \mathrm{pc\,cm}^{-6}.
\end{equation}

As far as we are aware, constraints on the emission measure have not been placed in CWB literature. To provide a comparison, we consider SN 1987A, which has an emission measure that is an order of magnitude smaller than the value we find for Apep \citep{Callingham2016}. The difference in emission measures is expected, since the photosphere of the WR stars also needs to be considered in Apep. 

A radio photon emitted from the shock region will have an equivalent path length $l$ that is roughly equal to the Str\"{o}mgen radius, which is $\approx$1\,pc for a WR star \citep{Vamvatira2016}. If we assume the ionized material is in a slab with a uniform density with the filling factor $f$=1, we find a value of the electron density $n_e = \sqrt{\mathrm{EM}/fl} \approx$ 900\,cm$^{-3}$. This is again an order of magnitude larger than the value found by \citet{Callingham2016}. As the electron temperature can be higher than the value we assume here, this is a lower limit.

As mentioned above, we show that IC cooling is needed to describe the entire spectrum of Apep. The characteristic break frequency we find is $\nu_{\mathrm{{break, IC}}}$=36$\pm$2\,GHz. The impact of IC cooling on the spectrum depends on the strength of the radiation field in the system. Since IC cooling is impacting a relatively low frequency, this implies a strong radiation field for a CWB \citep{Kellermann1969}. This is not unexpected, as there are two WR stars present in Apep.

\subsection{Temporal analysis}
\label{sec:temp_model}

The flux density of Apep shows significant variation over time at the low frequencies. At the high frequencies, the flux density remains approximately constant over a 10-year period. The low-frequency variation can be explained with the variation of free-free absorption as the WCR moves through the winds launched by the WR stars.
Spherically symmetric winds cannot replicate all of the structure we see in the lightcurve of Apep, as shown in Section\,\ref{sec:temp_model}. Although it can explain certain features, it does not explain the decrease in the flux density at 1.4\,GHz we observe in our targeted observing campaign from 2017 to 2021. The spherical winds model also consistently overestimates the flux density at 2.3\,GHz and underestimates the flux density at 843\,MHz. While a model with a very short period and zero eccentricity can fit the lightcurve, such a model is discounted as there is no evidence for a third body in the system from the infrared, optical, or VLBI observations of Apep. Although it is possible to describe some general structures in the lightcurve using the spherical wind model, it is a poor fit.

Since spherically symmetric winds cannot reproduce the lightcurve completely, we developed a model that has an anisotropic wind. In this model, the wind from one of the WR stars is described by a vertical Gaussian, which is able to fit the data significantly better than the spherical winds model, with a difference in the log of the Bayesian evidence values of $\sim$22. 

The lightcurve of Apep is not symmetric around the peak we observe around 2017, which implies that the slow wind and the orbit of the WCR are not aligned. The parameters of the anisotropic model show this as well. If the anisotropic wind is driven by near-critical rotation of the WR star, this means that the rotation axis of the star and the orbital axis of the system are not aligned.
We can also break the degeneracy in the inclination of the orbit with the anisotropic model because of the misalignment between the orbit and the slow wind. Although it is possible to fit a model to the data with a positive inclination of the orbit to the data, such a model would require an inclination of the wind of $\sim$70\degree. This is not possible, since we measure a fast wind along the line of sight and a slower wind in the plane of the sky. The slow wind therefore needs to be at least close to the plane of the sky, and cannot be almost perpendicular to it.

The best-fitting parameters of the anisotropic wind model agree with the orbital parameters found by fitting the infrared data \citep{han2020}. With the spherical wind model, it is not possible to find a solution that describes all of the structure in the radio lightcurve while also being consistent with the orbital elements derived from the infrared observations.
Therefore, we suggest that our modelling of the radio lightcurve of Apep shows that there is evidence in the radio lightcurve for an anisotropic wind launched by one of the WR stars in Apep. If this model is correct, we predict a $\sim$25\% decrease in the flux density at 1.4\,GHz over the next 5 years. By 2035, it will be completely obscured at the low frequencies.
\citet{Callingham2019} already suggested that Apep could have an anisotropic wind structure based on mid-infrared observations. We also independently suggest that there is evidence from the radio lightcurve that Apep has an anisotropic wind structure. %However, we do note that the anisotropic wind model predicts a substantial phase change in the radio lightcurve after the end of our measurements, which needs to be confirmed over the next $\sim$5 years. 

The parameters of the best fit of the anisotropic model show that the density of the wind is reasonably high. If Apep had a spherical wind instead, with the same density, it would be almost completely obscured at the low frequencies. This could possibly explain why Apep is so bright for a CWB, as it is less obscured than other CWBs that do have spherical winds.

Finally, we note that a potential different explanation for the variation we see in the lightcurve of Apep is ``micro-variability'' caused by variable shock conditions in the WCR. \citet{SetiaGunawan2000} find evidence of micro-variability in the lightcurve of WR\,146. They explain this varibaility as being caused by clumps entering the WCR, which result in variations in the FFA opacity. The effects they observe cause at most 10\,mJy of variation. We observe a variability on the order of 50\,mJy at 1.4\,GHz. Furthermore, \citet{SetiaGunawan2000} state that micro-variability could be caused by irregularities in the mass inflow. These irregularities change the non-thermal synchrotron emission, which has to result in a variation of the entire spectrum. However, we see no significant flux density variation at 5.5, 9, or 20\,GHz, so it is unlikely that the variation we observe is caused by micro-variability.

\section{Conclusions}
\label{sec:concl}

We have analysed radio data on the brightest and most luminous non-thermal emitting CWB Apep. The data span from 0.2 to 20\,GHz and a time baseline of 27 years at 1.4\,GHz.

Our study of the radio spectrum of Apep has revealed that the emission is best described by synchrotron emission combined with FFA and inverse-Compton cooling. The spectral index of the synchrotron emission is consistent with other non-thermal emitting CWBs at -0.52$\pm$0.02. The FFA turnover frequency is 0.54$\pm$0.01\,GHz, which leads to a rough estimate that the electron density $n_e$ $\gtrsim 900$\,cm$^{-3}$. The characteristic IC cooling frequency is 36\,$\pm$\,2\,GHz. This indicates that there is a strong radiation field in Apep, consistent with Apep being composed of two WR stars.

Radio monitoring observations of Apep over a 33-year period show that the flux density of Apep is not constant. The time variability of the flux density of Apep can be modelled as variation in the optical depth of free-free absorption as the WCR moves through the winds of the binary. A model with spherical winds, such as that developed by \citet{williams1990}, describes some but not all of the structure visible in the lightcurve. The spherical winds model consistently overestimates the flux density at 2.3\,GHz and underestimates it at 843\,MHz. Furthermore, it does not predict the decrease in flux density at 1.4\,GHz that we observe in our targeted observing campaign since 2017.

We develop an anisotropic wind model that is better able to replicate the radio lightcurve of Apep. This model can recreate the asymmetry in the lightcurve because the slow wind and the orbit are not aligned. This implies that the rotation axis of the WR star launching the anisotropic wind and the orbital axis of the system are not aligned.
The anisotropic mass-loss model also derives orbital elements that are consistent with those derived from dust plume and binary detected in mid-infrared data \citep{han2020}. We believe this is a significant result since this provides independent evidence of anisotropic mass-loss first suggested by infrared data on Apep. If our anisotropic mass-loss model is correct, we predict a $\sim$25\% decrease in the flux density at 1.4\,GHz over the next 5 years. This model also suggests that one of the reasons Apep is so radio bright is because the amount of FFA occurring is relatively low compared to other CWBs. 

The model for the anisotropic wind we developed in this work is a first look at attempting to explain radio variation with non-spherical winds in a CWB. There are several assumptions in the model that could be changed if there is sufficient evidence to justify the change. For example, the power-law component of the density distribution of the slow wind is assumed to have an index of -3.5, but could take another value. However, changing this index does not seem to have a strong impact on the results from our tests. Furthermore, the model could also be improved by adding a fast polar wind. We ignored the fast polar wind in our model because it has a very low density compared to the slow equatorial wind. This implies the impact of FFA in that region should be minimal. However, in more sophisticated models this will have to be taken into account and likely will mean higher frequency data ($\gtrsim$5\,GHz) would see variation as the WCR enters and exits the fast polar wind.

Finally, if the optical depth at the shock region is indeed changing as much as our model suggests, there should be variation in the X-ray emission as well. A monitoring campaign of Apep at these X-ray frequencies would also be an independent test of our anisotropic wind model.

\section*{Acknowledgements}

J.~R.~C. thanks the Nederlandse Organisatie voor Wetenschappelijk Onderzoek (NWO) for support via the Talent Programme Veni grant.
The Australia Telescope Compact Array is part of the Australia Telescope National Facility which is funded by the Australian Government for operation as a National Facility managed by CSIRO. We acknowledge the Gomeroi people as the traditional owners of the Observatory site.
The Giant Metrewave Radio Telescope (GMRT) is run by the National Centre for Radio Astrophysics of the Tata Institute of Fundamental Research (NCRA-TIFR). B.M. acknowledges support from the Spanish Ministerio de Econom\'ia y Competitividad (MINECO) under grant AYA2016-76012-C3-1-P and from the Spanish Ministerio de Ciencia e Innovaci\'on under grants PID2019-105510GB-C31 and CEX2019-000918-M of ICCUB (Unidad de Excelencia ``Mar\'ia de Maeztu'' 2020-2023).
We thank Y.~Han for his help with the ATCA observations in September 2019.
This research made use of NASA's Astrophysics Data System, the \textsc{IPython} package \citep{PER-GRA:2007}; \textsc{SciPy} \citep{scipy}; \textsc{Matplotlib}, a \textsc{Python} library for publication quality graphics \citep{Hunter:2007}; \textsc{Astropy}, a community-developed core \textsc{Python} package for astronomy \citep{2013A&A...558A..33A}; and \textsc{NumPy} \citep{van2011numpy}. 

\section*{Data availability}
The data underlying this article are available in the Australia Telescope Online Archive at \url{https://atoa.atnf.csiro.au/}, and in the GMRT Online Archive at \url{https://naps.ncra.tifr.res.in/goa/data/}. The models of the variation in free-free absorption are available online on GitHub at \url{https://github.com/SBloot/FFA-models}.

%%%%%%%%%%%%%%%%%%%%%%%%%%%%%%%%%%%%%%%%%%%%%%%%%%

%%%%%%%%%%%%%%%%%%%% REFERENCES %%%%%%%%%%%%%%%%%%

\bibliographystyle{mnras}
\bibliography{mybib_stars_paper.bib}

\begin{thebibliography}{}
\makeatletter
\relax
\def\mn@urlcharsother{\let\do\@makeother \do\$\do\&\do\#\do\^\do\_\do\%\do\~}
\def\mn@doi{\begingroup\mn@urlcharsother \@ifnextchar [ {\mn@doi@}
  {\mn@doi@[]}}
\def\mn@doi@[#1]#2{\def\@tempa{#1}\ifx\@tempa\@empty \href
  {http://dx.doi.org/#2} {doi:#2}\else \href {http://dx.doi.org/#2} {#1}\fi
  \endgroup}
\def\mn@eprint#1#2{\mn@eprint@#1:#2::\@nil}
\def\mn@eprint@arXiv#1{\href {http://arxiv.org/abs/#1} {{\tt arXiv:#1}}}
\def\mn@eprint@dblp#1{\href {http://dblp.uni-trier.de/rec/bibtex/#1.xml}
  {dblp:#1}}
\def\mn@eprint@#1:#2:#3:#4\@nil{\def\@tempa {#1}\def\@tempb {#2}\def\@tempc
  {#3}\ifx \@tempc \@empty \let \@tempc \@tempb \let \@tempb \@tempa \fi \ifx
  \@tempb \@empty \def\@tempb {arXiv}\fi \@ifundefined
  {mn@eprint@\@tempb}{\@tempb:\@tempc}{\expandafter \expandafter \csname
  mn@eprint@\@tempb\endcsname \expandafter{\@tempc}}}

\bibitem[\protect\citeauthoryear{{Astropy Collaboration} et~al.,}{{Astropy
  Collaboration} et~al.}{2013}]{2013A&A...558A..33A}
{Astropy Collaboration} et~al., 2013, \mn@doi [\aap]
  {10.1051/0004-6361/201322068}, \href
  {http://adsabs.harvard.edu/abs/2013A%26A...558A..33A} {558, A33}

\bibitem[\protect\citeauthoryear{{Axford}, {Leer}  \& {Skadron}}{{Axford}
  et~al.}{1977}]{Axford1977}
{Axford} W.~I.,  {Leer} E.,   {Skadron} G.,  1977, in International Cosmic Ray
  Conference. p.~132

\bibitem[\protect\citeauthoryear{{Bell}}{{Bell}}{1978}]{Bell1978}
{Bell} A.~R.,  1978, \mn@doi [\mnras] {10.1093/mnras/182.2.147}, \href
  {https://ui.adsabs.harvard.edu/abs/1978MNRAS.182..147B} {182, 147}

\bibitem[\protect\citeauthoryear{{Benaglia}, {De Becker}, {Ishwara-Chandra},
  {Intema}  \& {Isequilla}}{{Benaglia} et~al.}{2020}]{Benaglia2020}
{Benaglia} P.,  {De Becker} M.,  {Ishwara-Chandra} C.~H.,  {Intema} H.~T.,
  {Isequilla} N.~L.,  2020, \mn@doi [\pasa] {10.1017/pasa.2020.21}, \href
  {https://ui.adsabs.harvard.edu/abs/2020PASA...37...30B} {37, e030}

\bibitem[\protect\citeauthoryear{{Bjorkman}}{{Bjorkman}}{1997}]{Bjorkman1997}
{Bjorkman} J.~E.,  1997, {Circumstellar Disks}.
Springer, p.~239, \mn@doi{10.1007/BFb0113487}

\bibitem[\protect\citeauthoryear{{Blandford} \& {Ostriker}}{{Blandford} \&
  {Ostriker}}{1978}]{Blandford1978}
{Blandford} R.~D.,  {Ostriker} J.~P.,  1978, \mn@doi [\apjl] {10.1086/182658},
  \href {https://ui.adsabs.harvard.edu/abs/1978ApJ...221L..29B} {221, L29}

\bibitem[\protect\citeauthoryear{{Callingham} et~al.,}{{Callingham}
  et~al.}{2015}]{Callingham2015}
{Callingham} J.~R.,  et~al., 2015, \mn@doi [\apj]
  {10.1088/0004-637X/809/2/168}, \href
  {http://adsabs.harvard.edu/abs/2015ApJ...809..168C} {809, 168}

\bibitem[\protect\citeauthoryear{{Callingham} et~al.,}{{Callingham}
  et~al.}{2016}]{Callingham2016}
{Callingham} J.~R.,  et~al., 2016, \mn@doi [\mnras] {10.1093/mnras/stw1489},
  \href {https://ui.adsabs.harvard.edu/abs/2016MNRAS.462..290C} {462, 290}

\bibitem[\protect\citeauthoryear{{Callingham}, {Tuthill}, {Pope}, {Williams},
  {Crowther}, {Edwards}, {Norris}  \& {Kedziora-Chudczer}}{{Callingham}
  et~al.}{2019}]{Callingham2019}
{Callingham} J.~R.,  {Tuthill} P.~G.,  {Pope} B.~J.~S.,  {Williams} P.~M.,
  {Crowther} P.~A.,  {Edwards} M.,  {Norris} B.,   {Kedziora-Chudczer} L.,
  2019, \mn@doi [Nature Astronomy] {10.1038/s41550-018-0617-7}, \href
  {https://ui.adsabs.harvard.edu/abs/2019NatAs...3...82C} {3, 82}

\bibitem[\protect\citeauthoryear{{Callingham}, {Crowther}, {Williams},
  {Tuthill}, {Han}, {Pope}  \& {Marcote}}{{Callingham}
  et~al.}{2020}]{Callingham2020}
{Callingham} J.~R.,  {Crowther} P.~A.,  {Williams} P.~M.,  {Tuthill} P.~G.,
  {Han} Y.,  {Pope} B.~J.~S.,   {Marcote} B.,  2020, \mn@doi [\mnras]
  {10.1093/mnras/staa1244}, \href
  {https://ui.adsabs.harvard.edu/abs/2020MNRAS.495.3323C} {495, 3323}

\bibitem[\protect\citeauthoryear{{Carciofi} \& {Bjorkman}}{{Carciofi} \&
  {Bjorkman}}{2006}]{Carciofi2006}
{Carciofi} A.~C.,  {Bjorkman} J.~E.,  2006, \mn@doi [\apj] {10.1086/499483},
  \href {https://ui.adsabs.harvard.edu/abs/2006ApJ...639.1081C} {639, 1081}

\bibitem[\protect\citeauthoryear{{Chen}, {Guo}, {Yu}  \& {Takata}}{{Chen}
  et~al.}{2021}]{Chen2021}
{Chen} A.~M.,  {Guo} Y.~D.,  {Yu} Y.~W.,   {Takata} J.,  2021, arXiv e-prints,
  \href {https://ui.adsabs.harvard.edu/abs/2021arXiv210610445C} {p.
  arXiv:2106.10445}

\bibitem[\protect\citeauthoryear{{Chini}, {Hoffmeister}, {Nasseri}, {Stahl}  \&
  {Zinnecker}}{{Chini} et~al.}{2012}]{Chini2012}
{Chini} R.,  {Hoffmeister} V.~H.,  {Nasseri} A.,  {Stahl} O.,   {Zinnecker} H.,
   2012, \mn@doi [\mnras] {10.1111/j.1365-2966.2012.21317.x}, \href
  {https://ui.adsabs.harvard.edu/abs/2012MNRAS.424.1925C} {424, 1925}

\bibitem[\protect\citeauthoryear{{Crowther}}{{Crowther}}{2007}]{Crowther2007}
{Crowther} P.~A.,  2007, \mn@doi [\araa]
  {10.1146/annurev.astro.45.051806.110615}, \href
  {https://ui.adsabs.harvard.edu/abs/2007ARA&A..45..177C} {45, 177}

\bibitem[\protect\citeauthoryear{{De Becker} \& {Raucq}}{{De Becker} \&
  {Raucq}}{2013}]{debecker2013}
{De Becker} M.,  {Raucq} F.,  2013, \mn@doi [\aap]
  {10.1051/0004-6361/201322074}, \href
  {http://adsabs.harvard.edu/abs/2013A%26A...558A..28D} {558, A28}

\bibitem[\protect\citeauthoryear{{Detmers}, {Langer}, {Podsiadlowski}  \&
  {Izzard}}{{Detmers} et~al.}{2008}]{Detmers2008}
{Detmers} R.~G.,  {Langer} N.,  {Podsiadlowski} P.,   {Izzard} R.~G.,  2008,
  \mn@doi [\aap] {10.1051/0004-6361:200809371}, \href
  {http://adsabs.harvard.edu/abs/2008A%26A...484..831D} {484, 831}

\bibitem[\protect\citeauthoryear{{Dougherty}, {Pittard}, {Kasian}, {Coker},
  {Williams}  \& {Lloyd}}{{Dougherty} et~al.}{2003}]{Dougherty2003}
{Dougherty} S.~M.,  {Pittard} J.~M.,  {Kasian} L.,  {Coker} R.~F.,  {Williams}
  P.~M.,   {Lloyd} H.~M.,  2003, \mn@doi [\aap] {10.1051/0004-6361:20031048},
  \href {https://ui.adsabs.harvard.edu/abs/2003A&A...409..217D} {409, 217}

\bibitem[\protect\citeauthoryear{{Feroz}, {Hobson}, {Cameron}  \&
  {Pettitt}}{{Feroz} et~al.}{2013}]{Feroz2013}
{Feroz} F.,  {Hobson} M.~P.,  {Cameron} E.,   {Pettitt} A.~N.,  2013,
  arXiv:1306.2144, \href {http://adsabs.harvard.edu/abs/2013arXiv1306.2144F} {}

\bibitem[\protect\citeauthoryear{{Foreman-Mackey}, {Hogg}, {Lang}  \&
  {Goodman}}{{Foreman-Mackey} et~al.}{2013}]{ForemanMackey2013}
{Foreman-Mackey} D.,  {Hogg} D.~W.,  {Lang} D.,   {Goodman} J.,  2013, \mn@doi
  [\pasp] {10.1086/670067}, \href
  {http://adsabs.harvard.edu/abs/2013PASP..125..306F} {125, 306}

\bibitem[\protect\citeauthoryear{Goodman \& Weare}{Goodman \&
  Weare}{2010}]{Goodman2010}
Goodman J.,  Weare J.,  2010, Communications in Applied Mathematics and
  Computational Science, 5, 65

\bibitem[\protect\citeauthoryear{{Han} et~al.,}{{Han} et~al.}{2020}]{han2020}
{Han} Y.,  et~al., 2020, \mn@doi [\mnras] {10.1093/mnras/staa2349}, \href
  {https://ui.adsabs.harvard.edu/abs/2020MNRAS.498.5604H} {498, 5604}

\bibitem[\protect\citeauthoryear{{Hancock}, {Murphy}, {Gaensler}, {Hopkins}  \&
  {Curran}}{{Hancock} et~al.}{2012}]{aegean_2}
{Hancock} P.~J.,  {Murphy} T.,  {Gaensler} B.~M.,  {Hopkins} A.,   {Curran}
  J.~R.,  2012, \mn@doi [\mnras] {10.1111/j.1365-2966.2012.20768.x}, \href
  {https://ui.adsabs.harvard.edu/abs/2012MNRAS.422.1812H} {422, 1812}

\bibitem[\protect\citeauthoryear{{Hancock}, {Trott}  \&
  {Hurley-Walker}}{{Hancock} et~al.}{2018}]{aegean}
{Hancock} P.~J.,  {Trott} C.~M.,   {Hurley-Walker} N.,  2018, \mn@doi [\pasa]
  {10.1017/pasa.2018.3}, \href
  {https://ui.adsabs.harvard.edu/abs/2018PASA...35...11H} {35, e011}

\bibitem[\protect\citeauthoryear{{Hornby} \& {Williams}}{{Hornby} \&
  {Williams}}{1966}]{Hornby1966}
{Hornby} J.~M.,  {Williams} P.~J.~S.,  1966, \mn@doi [\mnras]
  {10.1093/mnras/131.2.237}, \href
  {https://ui.adsabs.harvard.edu/abs/1966MNRAS.131..237H} {131, 237}

\bibitem[\protect\citeauthoryear{Hunter}{Hunter}{2007}]{Hunter:2007}
Hunter J.~D.,  2007, Computing In Science \& Engineering, 9, 90

\bibitem[\protect\citeauthoryear{Jones, Oliphant, Peterson  \& Others}{Jones
  et~al.}{2001}]{scipy}
Jones E.,  Oliphant T.,  Peterson P.,   Others 2001, {SciPy}: Open source
  scientific tools for Python, \url {http://www.scipy.org/}

\bibitem[\protect\citeauthoryear{{Kale} \& {Ishwara-Chandra}}{{Kale} \&
  {Ishwara-Chandra}}{2021}]{kale2021}
{Kale} R.,  {Ishwara-Chandra} C.~H.,  2021, \mn@doi [Experimental Astronomy]
  {10.1007/s10686-020-09677-6}, \href
  {https://ui.adsabs.harvard.edu/abs/2021ExA....51...95K} {51, 95}

\bibitem[\protect\citeauthoryear{Kass \& Raftery}{Kass \&
  Raftery}{1995}]{Kass1995}
Kass R.~E.,  Raftery A.~E.,  1995, Journal of the American Statistical
  Association, 90, 773

\bibitem[\protect\citeauthoryear{{Kellermann} \& {Pauliny-Toth}}{{Kellermann}
  \& {Pauliny-Toth}}{1969}]{Kellermann1969}
{Kellermann} K.~I.,  {Pauliny-Toth} I.~I.~K.,  1969, \mn@doi [\apjl]
  {10.1086/180305}, \href
  {https://ui.adsabs.harvard.edu/abs/1969ApJ...155L..71K} {155, L71}

\bibitem[\protect\citeauthoryear{{Kennedy}, {Dougherty}, {Fink}  \&
  {Williams}}{{Kennedy} et~al.}{2010}]{Kennedy2010}
{Kennedy} M.,  {Dougherty} S.~M.,  {Fink} A.,   {Williams} P.~M.,  2010,
  \mn@doi [\apj] {10.1088/0004-637X/709/2/632}, \href
  {https://ui.adsabs.harvard.edu/abs/2010ApJ...709..632K} {709, 632}

\bibitem[\protect\citeauthoryear{{Komissarov} \& {Gubanov}}{{Komissarov} \&
  {Gubanov}}{1994}]{Komissarov1994}
{Komissarov} S.~S.,  {Gubanov} A.~G.,  1994, \aap, \href
  {http://adsabs.harvard.edu/abs/1994A%26A...285...27K} {285, 27}

\bibitem[\protect\citeauthoryear{{Lamers}, {Maeder}, {Schmutz}  \&
  {Cassinelli}}{{Lamers} et~al.}{1991}]{Lamers1991}
{Lamers} H.~J.~G.~L.~M.,  {Maeder} A.,  {Schmutz} W.,   {Cassinelli} J.~P.,
  1991, \mn@doi [\apj] {10.1086/169717}, \href
  {https://ui.adsabs.harvard.edu/abs/1991ApJ...368..538L} {368, 538}

\bibitem[\protect\citeauthoryear{{Leitherer}, {Chapman}  \&
  {Koribalski}}{{Leitherer} et~al.}{1997}]{Leitherer1997}
{Leitherer} C.,  {Chapman} J.~M.,   {Koribalski} B.,  1997, \mn@doi [\apj]
  {10.1086/304096}, \href {http://adsabs.harvard.edu/abs/1997ApJ...481..898L}
  {481, 898}

\bibitem[\protect\citeauthoryear{MacFadyen \& Woosley}{MacFadyen \&
  Woosley}{1999}]{MacFayden1999}
MacFadyen A.~I.,  Woosley S.~E.,  1999, \mn@doi [\apj] {10.1086/307790}, 524,
  262

\bibitem[\protect\citeauthoryear{{MacFadyen}, {Woosley}  \&
  {Heger}}{{MacFadyen} et~al.}{2001}]{MacFayden2001}
{MacFadyen} A.~I.,  {Woosley} S.~E.,   {Heger} A.,  2001, \mn@doi [\apj]
  {10.1086/319698}, \href
  {https://ui.adsabs.harvard.edu/abs/2001ApJ...550..410M} {550, 410}

\bibitem[\protect\citeauthoryear{{Marcote}, {Callingham}, {De Becker},
  {Edwards}, {Han}, {Schulz}, {Stevens}  \& {Tuthill}}{{Marcote}
  et~al.}{2021}]{marcote2021}
{Marcote} B.,  {Callingham} J.~R.,  {De Becker} M.,  {Edwards} P.~G.,  {Han}
  Y.,  {Schulz} R.,  {Stevens} J.,   {Tuthill} P.~G.,  2021, \mn@doi [\mnras]
  {10.1093/mnras/staa3863}, \href
  {https://ui.adsabs.harvard.edu/abs/2021MNRAS.501.2478M} {501, 2478}

\bibitem[\protect\citeauthoryear{McConnell et~al.,}{McConnell
  et~al.}{2020}]{RACS_2020}
McConnell D.,  et~al., 2020, \mn@doi [Publications of the Astronomical Society
  of Australia] {10.1017/pasa.2020.41}, 37, e048

\bibitem[\protect\citeauthoryear{{Meynet} \& {Maeder}}{{Meynet} \&
  {Maeder}}{2005}]{Meynet2005}
{Meynet} G.,  {Maeder} A.,  2005, \mn@doi [A\&A] {10.1051/0004-6361:20047106},
  429, 581

\bibitem[\protect\citeauthoryear{{Mezger} \& {Henderson}}{{Mezger} \&
  {Henderson}}{1967}]{Mezger1967}
{Mezger} P.~G.,  {Henderson} A.~P.,  1967, \mn@doi [\apj] {10.1086/149030},
  \href {http://adsabs.harvard.edu/abs/1967ApJ...147..471M} {147, 471}

\bibitem[\protect\citeauthoryear{{Monnier}, {Greenhill}, {Tuthill}  \&
  {Danchi}}{{Monnier} et~al.}{2002}]{Monnier2002}
{Monnier} J.~D.,  {Greenhill} L.~J.,  {Tuthill} P.~G.,   {Danchi} W.~C.,  2002,
  \mn@doi [\apj] {10.1086/337961}, \href
  {http://adsabs.harvard.edu/abs/2002ApJ...566..399M} {566, 399}

\bibitem[\protect\citeauthoryear{{Murphy}, {Mauch}, {Green}, {Hunstead},
  {Piestrzynska}, {Kels}  \& {Sztajer}}{{Murphy} et~al.}{2007}]{MOST}
{Murphy} T.,  {Mauch} T.,  {Green} A.,  {Hunstead} R.~W.,  {Piestrzynska} B.,
  {Kels} A.~P.,   {Sztajer} P.,  2007, \mn@doi [\mnras]
  {10.1111/j.1365-2966.2007.12379.x}, \href
  {https://ui.adsabs.harvard.edu/abs/2007MNRAS.382..382M} {382, 382}

\bibitem[\protect\citeauthoryear{Nelder \& Mead}{Nelder \&
  Mead}{1965}]{Nelder1965}
Nelder J.~A.,  Mead R.,  1965, The Computer Journal, 7, 308

\bibitem[\protect\citeauthoryear{P\'erez \& Granger}{P\'erez \&
  Granger}{2007}]{PER-GRA:2007}
P\'erez F.,  Granger B.~E.,  2007, \mn@doi [Computing in Science and
  Engineering] {10.1109/MCSE.2007.53}, 9, 21

\bibitem[\protect\citeauthoryear{{Perley} \& {Butler}}{{Perley} \&
  {Butler}}{2013}]{Perley2013}
{Perley} R.~A.,  {Butler} B.~J.,  2013, \mn@doi [\apjs]
  {10.1088/0067-0049/204/2/19}, \href
  {https://ui.adsabs.harvard.edu/abs/2013ApJS..204...19P} {204, 19}

\bibitem[\protect\citeauthoryear{{Pittard} \& {Dougherty}}{{Pittard} \&
  {Dougherty}}{2006}]{Pittard2_2006}
{Pittard} J.~M.,  {Dougherty} S.~M.,  2006, \mn@doi [\mnras]
  {10.1111/j.1365-2966.2006.10888.x}, \href
  {https://ui.adsabs.harvard.edu/abs/2006MNRAS.372..801P} {372, 801}

\bibitem[\protect\citeauthoryear{{Pittard}, {Dougherty}, {Coker}, {O'Connor}
  \& {Bolingbroke}}{{Pittard} et~al.}{2006}]{Pittard2006}
{Pittard} J.~M.,  {Dougherty} S.~M.,  {Coker} R.~F.,  {O'Connor} E.,
  {Bolingbroke} N.~J.,  2006, \mn@doi [\aap] {10.1051/0004-6361:20053649},
  \href {https://ui.adsabs.harvard.edu/abs/2006A&A...446.1001P} {446, 1001}

\bibitem[\protect\citeauthoryear{Rybicki \& Lightman}{Rybicki \&
  Lightman}{1979}]{Rybicki_Lightman}
Rybicki G.~B.,  Lightman A.~P.,  1979, Radiative processes in astrophysics.
John Wiley \& Sons

\bibitem[\protect\citeauthoryear{Sana et~al.,}{Sana et~al.}{2014}]{Sana2014}
Sana H.,  et~al., 2014, \mn@doi [The Astrophysical Journal Supplement Series]
  {10.1088/0067-0049/215/1/15}, 215, 15

\bibitem[\protect\citeauthoryear{{Scaife} \& {Heald}}{{Scaife} \&
  {Heald}}{2012}]{Scaife2012}
{Scaife} A.~M.~M.,  {Heald} G.~H.,  2012, \mn@doi [\mnras]
  {10.1111/j.1745-3933.2012.01251.x}, \href
  {http://adsabs.harvard.edu/abs/2012MNRAS.423L..30S} {423, L30}

\bibitem[\protect\citeauthoryear{{Setia Gunawan}, {de Bruyn}, {van der Hucht}
  \& {Williams}}{{Setia Gunawan} et~al.}{2000}]{SetiaGunawan2000}
{Setia Gunawan} D.~Y.~A.,  {de Bruyn} A.~G.,  {van der Hucht} K.~A.,
  {Williams} P.~M.,  2000, \aap, \href
  {https://ui.adsabs.harvard.edu/abs/2000A&A...356..676S} {356, 676}

\bibitem[\protect\citeauthoryear{Thompson}{Thompson}{1994}]{Thompson1994}
Thompson C.,  1994, \mn@doi [\mnras] {10.1093/mnras/270.3.480}, 270, 480

\bibitem[\protect\citeauthoryear{{Tingay} \& {de Kool}}{{Tingay} \& {de
  Kool}}{2003}]{Tingay2003}
{Tingay} S.~J.,  {de Kool} M.,  2003, \mn@doi [\aj] {10.1086/376600}, \href
  {http://adsabs.harvard.edu/abs/2003AJ....126..723T} {126, 723}

\bibitem[\protect\citeauthoryear{{Vamvatira-Nakou}, {Hutsem\'ekers}, {Royer},
  {Waelkens}, {Groenewegen}  \& {Barlow}}{{Vamvatira-Nakou}
  et~al.}{2016}]{Vamvatira2016}
{Vamvatira-Nakou} C.,  {Hutsem\'ekers} D.,  {Royer} P.,  {Waelkens} C.,
  {Groenewegen} M. A.~T.,   {Barlow} M.~J.,  2016, \mn@doi [A\&A]
  {10.1051/0004-6361/201527667}, 588, A92

\bibitem[\protect\citeauthoryear{Van Der~Walt, Colbert  \& Varoquaux}{Van
  Der~Walt et~al.}{2011}]{van2011numpy}
Van Der~Walt S.,  Colbert S.~C.,   Varoquaux G.,  2011, Computing in Science \&
  Engineering, 13, 22

\bibitem[\protect\citeauthoryear{{Waters}, {van den Heuvel}, {Taylor}, {Habets}
   \& {Persi}}{{Waters} et~al.}{1988}]{Waters1988}
{Waters} L.~B.~F.~M.,  {van den Heuvel} E.~P.~J.,  {Taylor} A.~R.,  {Habets}
  G.~M.~H.~J.,   {Persi} P.,  1988, \aap, \href
  {https://ui.adsabs.harvard.edu/abs/1988A&A...198..200W} {198, 200}

\bibitem[\protect\citeauthoryear{{Williams}, {van der Hucht}, {Pollock},
  {Florkowski}, {van der Woerd}  \& {Wamsteker}}{{Williams}
  et~al.}{1990}]{williams1990}
{Williams} P.~M.,  {van der Hucht} K.~A.,  {Pollock} A.~M.~T.,  {Florkowski}
  D.~R.,  {van der Woerd} H.,   {Wamsteker} W.~M.,  1990, \mnras, \href
  {https://ui.adsabs.harvard.edu/abs/1990MNRAS.243..662W} {243, 662}

\bibitem[\protect\citeauthoryear{{Wilson} et~al.,}{{Wilson}
  et~al.}{2011}]{Wilson2011}
{Wilson} W.~E.,  et~al., 2011, \mn@doi [\mnras]
  {10.1111/j.1365-2966.2011.19054.x}, \href
  {http://adsabs.harvard.edu/abs/2011MNRAS.416..832W} {416, 832}

\bibitem[\protect\citeauthoryear{{Woosley}}{{Woosley}}{1993}]{Woosley1993}
{Woosley} S.~E.,  1993, \mn@doi [\apj] {10.1086/172359}, \href
  {https://ui.adsabs.harvard.edu/abs/1993ApJ...405..273W} {405, 273}

\bibitem[\protect\citeauthoryear{{Woosley} \& {Heger}}{{Woosley} \&
  {Heger}}{2006}]{Woosley2006}
{Woosley} S.~E.,  {Heger} A.,  2006, \mn@doi [\apj] {10.1086/498500}, \href
  {http://adsabs.harvard.edu/abs/2006ApJ...637..914W} {637, 914}

\bibitem[\protect\citeauthoryear{{Wright} \& {Barlow}}{{Wright} \&
  {Barlow}}{1975}]{Wright1975}
{Wright} A.~E.,  {Barlow} M.~J.,  1975, \mnras, \href
  {http://adsabs.harvard.edu/abs/1975MNRAS.170...41W} {170, 41}

\bibitem[\protect\citeauthoryear{{de Mink}, {Langer}, {Izzard}, {Sana}  \& {de
  Koter}}{{de Mink} et~al.}{2013}]{DeMink2013}
{de Mink} S.~E.,  {Langer} N.,  {Izzard} R.~G.,  {Sana} H.,   {de Koter} A.,
  2013, \mn@doi [\apj] {10.1088/0004-637X/764/2/166}, \href
  {https://ui.adsabs.harvard.edu/abs/2013ApJ...764..166D} {764, 166}

\makeatother
\end{thebibliography}

\appendix

\section{Time Variability of Free-free Absorption} 

The free-free opacity is not constant over the orbit of the system. It depends on the amount of ionized medium between the source and the observer. In most CWBs, the two winds launched by the stars in the binary are both spherical. However, it has been theorised that one of the WR stars in Apep is launching an asymmetric wind. The derivation of the variation of optical depth for both models is presented in this appendix.

\subsection{Spherical winds model}
\label{app:williams}

A model for the general spherical winds case was derived by \citet{williams1990}. This model assumes that the winds are spherical, and that one of the stellar winds is dominating over the other.

\begin{figure}
    \centering
    \includegraphics[width=\columnwidth]{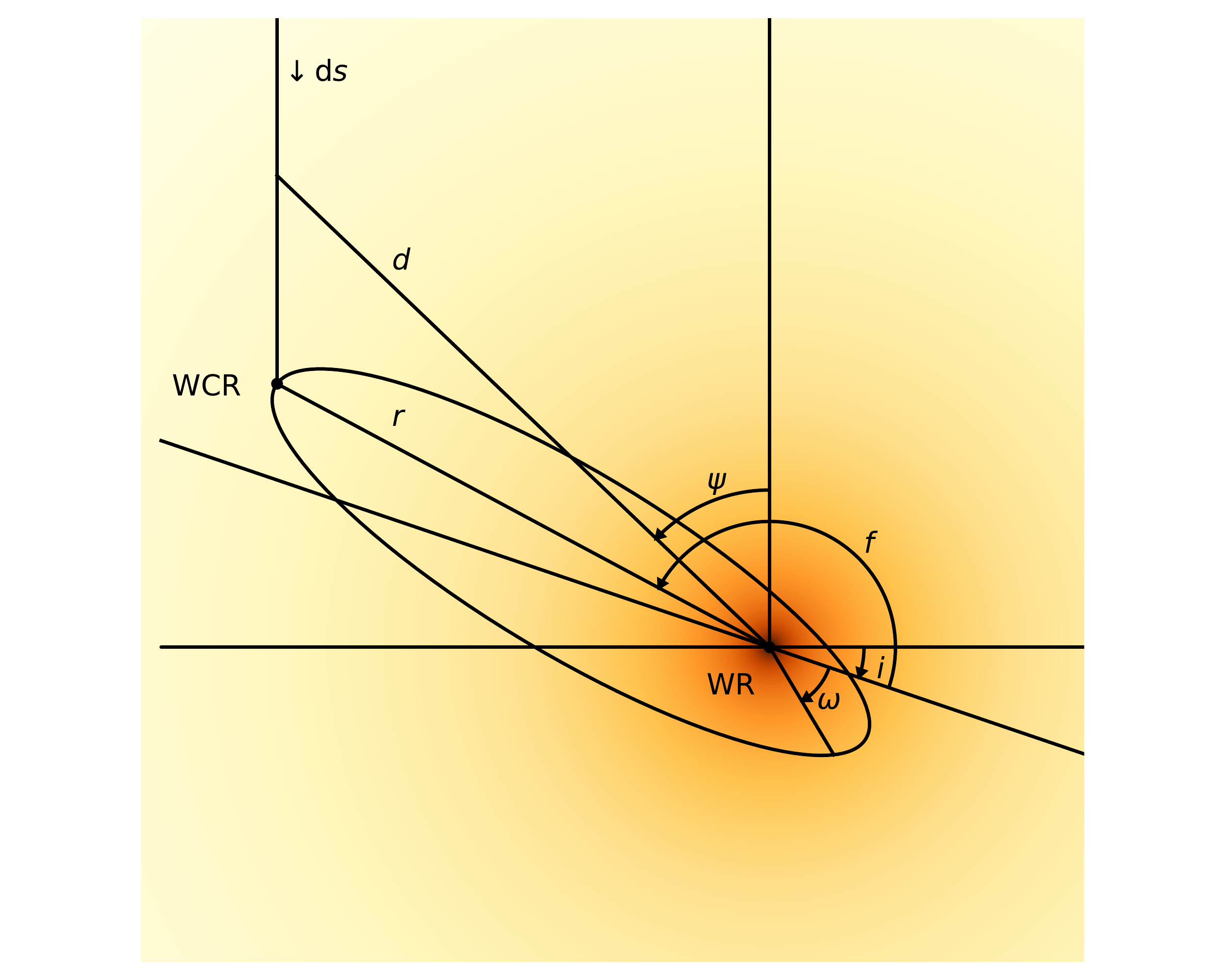}
    \caption{Geometry of the system seen from within the plane of the sky. The companion star is off the scale of the plot. The line of sight to the WCR is represented by the vertical line that ends at the location of the WCR. The colourmap represents the logarithm of the density of the wind. $i$ is the inclination of the orbit, $\omega$ is the argument of periastron, $f$ is the true anomaly of the WCR, $r$ is the separation between the WCR and the dominant WR star (indicated with WR in the figure) in units of the semi-major axis. $\psi$ is the angle over which we integrate.}
    \label{fig:orbit}
\end{figure}

Following \citet{williams1990}, we start with the standard free-free opacity, here given by Equation\,\ref{eq:ffa} in the main text, namely
\begin{equation*}
     \tau_{\mathrm{FFA}} \approx 8.24 \times 10^{-2} \nu^{-2.1} T_e^{-1.35} \int n_e^2 \mathrm{d}l,
\end{equation*}
where, as before, $\nu$ is the frequency in GHz, $T_e$ is the electron temperature in K, which we assume to be constant along the orbit, $n_e$ is the electron density in cm$^{-3}$ and $l$ is the distance through the source in pc.
As stated before, we assume that the situation at the location of the WCR is dominated by the wind of one of the WR stars. The geometry of the system is shown in Figure \ref{fig:orbit}.

The WR stars are always more than 40 $R_{*}$ apart, which means that the electron density in the wind $n_e \propto d^{-2}$ \citep[e.g.][]{Waters1988}. We can write $n_e^2(d) = n_e^2(a) d^{-4}$, where $n_e(a)$ is the number density of electrons in the wind at the semi-major axis $a$ and $d$ is the distance from the central WR star to the element $\mathrm{d}s$ in units of the semi-major axis. We can rewrite the expression for $\tau_{\mathrm{FFA}}$ assuming everything but $d$ is constant:
\begin{equation}
    \tau_{\mathrm{FFA}} = C_{\mathrm{FFA}} a^{-1} \int d^{-4} \mathrm{d}s,
    \label{eqn:a1}
\end{equation}
where $C_{\mathrm{FFA}} = \xi \left(\frac{\nu}{\nu_{\mathrm{ref}}}\right)^{ -2.1}$ \citep{Kennedy2010}. We use $\nu_{\mathrm{ref}}$=1.38\,GHz in this work. $\xi$ is a constant proportional to $n_e^2(a)$.

We can rewrite d$s$ as
\begin{equation*}
    \mathrm{d}s = r a \mathrm{cos}(\omega +f) \mathrm{cosec}^2(\psi) \mathrm{cosec}(i) \mathrm{d}\psi,
\end{equation*}
where $i$ is the inclination of the orbit, $\psi$ is the angle between indicated in Figure\,\ref{fig:orbit} and $r$ is the distance between the dominant WR star and the WCR, defined by
\begin{equation*}
    r = \frac{1-e^2}{(1+e \cos(f))}\frac{1-\eta^{1/2}}{1+\eta^{1/2}}.
\end{equation*}
In this equation, $e$ is the eccentricity and $f$ is the true anomaly. The momentum ratio $\eta$ is given by 
\begin{equation}
\eta=\frac{\dot{M}_{\mathrm{WC}} \mathrm{v}_{\infty, \mathrm{WC}}}{\dot{M}_{\mathrm{WN}} \mathrm{v}_{\infty, \mathrm{WN}}}.
\end{equation}

We can write $d$ as 
 \begin{equation*}
     d^2 = r^2 \mathrm{cos}^2(\omega + f) \{[\mathrm{cot}(\psi) + \mathrm{tan}(\omega+f)] \mathrm{cot}^2(i) + \mathrm{cosec}^2(\psi) \},
 \end{equation*}
where $\omega$ is the argument of periastron. We can now express the integral in Equation\,\ref{eqn:a1} in terms of these orbital parameters:

\begin{equation} 
\begin{split}
    a^{-1} & \int d^{-4} \mathrm{d}s = \\ 
    &\int (r^2 \mathrm{cos}^2(\omega + f)
    \{[\mathrm{cot}(\psi) + \mathrm{tan}(\omega+f)]\mathrm{cot}^2(i) +\\
    &\mathrm{cosec}^2(\psi) \})^{-2} r \mathrm{cos}(\omega +f) \mathrm{cosec}^2(\psi) \mathrm{cosec}(i) \mathrm{d}\psi.
\end{split}
\label{eqn:integral}
\end{equation}

Using trigonometric identities, Equation\,\ref{eqn:integral} can be rewritten as
\begin{equation*}
\begin{split}
    &a^{-1} \int d^{-4} \mathrm{d}s =\frac{r \mathrm{cos}(\omega+f)\mathrm{cosec}(i)}{r^4 \mathrm{cos}^4(\omega + f)} \\ 
    &\int \frac{\mathrm{cosec}^2(\psi)}{\{[\mathrm{cot}(\psi) + \mathrm{tan}(\omega+f)]^2+ \mathrm{tan}^2(i)\mathrm{cosec}^2(\psi)\} \mathrm{cot}^4(i)} \mathrm{d}\psi.
\end{split}
\end{equation*}

\begin{equation}
\label{int}
     =\frac{\mathrm{tan}^3(i)\mathrm{sec}(i)}{r^3 \mathrm{cos}^3(\omega + f)} \int \frac{\mathrm{cosec}^2(\psi)\mathrm{d}\psi}{[\mathrm{cot}(\psi) + \mathrm{tan}(\omega+f)]^2+ \mathrm{tan}^2(i)\mathrm{cosec}^2(\psi)}.
\end{equation}

By setting $t = \mathrm{tan}(\psi)$, this changes the integral into a form that has an analytical solution, namely:
\begin{equation*}
    \int \frac{t^2\mathrm{d}t}{\{ (1+\tan^2(i)) + 2 \tan(\omega+f) t + [\tan^2(\omega+f) + \tan^2(i)]t^2\}^2}.
\end{equation*}

We set the lower limit to $t$=0 and the upper limit to $t = \tan(\omega +f-3\pi/2)$. The solution to the integral is given by
\begin{equation}
\begin{split}
\label{eq:williams}
     a^{-1} \int d^{-4} \mathrm{d}s = \frac{\mathrm{sec}(i)}{2 \Delta r^3 \mathrm{cos}^3(\omega + f)}[\mathrm{sin}(\omega+f)\mathrm{cos}(\omega+f)\mathrm{tan}(i)+\\
     \frac{1+\mathrm{tan}^2(i)}{\sqrt{\Delta}}\mathrm{arctan}\left(\frac{-\sqrt{\Delta}}{\mathrm{tan}(\omega+f)\mathrm{tan}(i)}\right)],
    \end{split}
\end{equation}
where $$\Delta = 1+\mathrm{tan}^2(\omega+f)+\mathrm{tan}^2 (i).$$ \citet{williams1990} find the same result, except they do not have the $\Delta^{-1/2}$ term in the last term of their Equation\,A21. We believe this to be a typographical error, as the term does appear in other equations in the paper. We use Equation\,\ref{eq:williams} for our modelling.

Placing this result into Equation\,\ref{eqn:a1} provides us with the equation for $\tau_{\mathrm{FFA}}$ that can be used to fit for the orbital parameters of the system:

\begin{equation}
    \begin{split}
     \tau_{\mathrm{FFA}} = C_{\mathrm{FFA}}\frac{\mathrm{sec}(i)}{2 \Delta r^3 \mathrm{cos}^3(\omega + f)}[\mathrm{sin}(\omega+f)\mathrm{cos}(\omega+f)\mathrm{tan}(i)+\\
      \frac{1+\mathrm{tan}^2(i)}{\sqrt{\Delta}}\mathrm{arctan}\left(\frac{-\sqrt{\Delta}}{\mathrm{tan}(\omega+f)\mathrm{tan}(i)}\right)].
    \end{split}
\end{equation}
\begin{figure}
    \centering
    \includegraphics[width=\columnwidth]{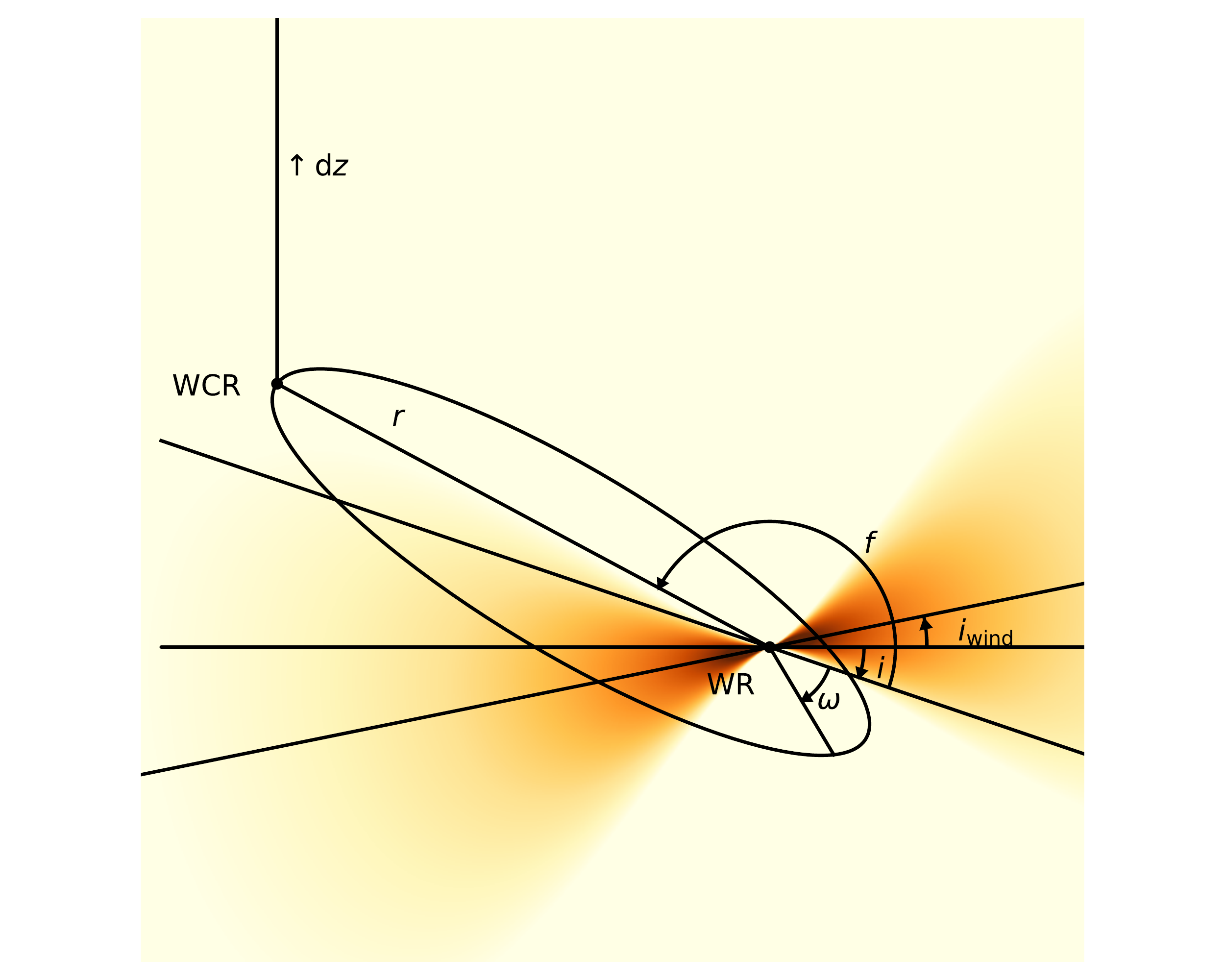}
    \caption{Geometry of the system with an anisotropic wind, seen from within the plane of the sky. The line of sight to the WCR is represented by the vertical line that ends at the location of the WCR. The colourmap represents the logarithm of the density of the wind. $i$ is the inclination of the orbit, $\omega$ is the argument of periastron, $f$ is the true anomaly of the WCR, $r$ is the separation between the WCR and the dominant WR star (indicated with WR in the figure) in units of the semi-major axis. $i_{\mathrm{wind}}$ is the inclination of the wind. Not shown in the figure is $\omega_{\mathrm{wind}}$, which is the angle over which the slow wind is rotated compared to the $x$-axis.}
    \label{fig:orbit_anisotropic}
\end{figure}

\subsection{Anisotropic wind model}
\label{app:disk}

As discussed in Section\,\ref{sec:intro}, one of the WR stars in Apep may be launching an anisotropic wind, consisting of a fast polar wind and a slow equatorial wind. The density of the slow wind can be described by a vertical Gaussian \citep[e.g.][]{Bjorkman1997,Carciofi2006}, as suggested by \citet{Chen2021}:
\begin{equation}
    \label{gaussian}
    n_e(d) = n_e(a) \left( {R^{\prime}} \right)^{-3.5} \exp \left( -\frac{z^{\prime 2}}{2 H^2(R^{\prime})} \right),
\end{equation}
where $H(R^{\prime}) = \frac{\mathrm{v}_s}{\mathrm{v}_c}{R^{\prime 3/2}}$ is the scale height of the slow wind, where $\mathrm{v}_s$ is the isothermal sound speed and $\mathrm{v}_c$ is the critical speed of the star. $R^{\prime}$ is the component of the position vector that lies in the plane of the slow wind and $z^{\prime}$ is the component of the position vector orthogonal to $R^{\prime}$, both in units of the semi-major axis.

The coordinates of the WCR at any point along its orbit are given by
\begin{equation} \label{eq:coords}
\begin{split}
x_{\mathrm{WCR}} & = r \cos(\omega - f) \cos(i), \\
y_{\mathrm{WCR}} & = r \sin(\omega - f), \\
z_{\mathrm{WCR}} & = r \cos(\omega - f) \sin(i), \\
\end{split}
\end{equation}
where $x_{\mathrm{WCR}}$ is the position of the WCR along the $f=0$ line
, $y_{\mathrm{WCR}}$ is the position along the $f=\pi/2$ line and $z_{\mathrm{WCR}}$ is the position of the WCR along the line orthogonal to both x and y, with positive z in our direction. All coordinates are in units of the semi-major axis.

The slow equatorial wind is not necessarily aligned with the orbit of the two stars. To account for this, we allow the wind to have an inclination $i_{\mathrm{wind}}$ and a rotation compared to the $x$-axis of $\omega_{\mathrm{wind}}$.
The density can be written as
\begin{equation*}
    n_e(d) = n_e(a) \left( \frac{R^{\prime}}{a} \right)^{-3.5} \exp \left( -\frac{z^{\prime 2}}{2 H^2(R^{\prime})} \right),
\end{equation*}
as before, with $R^{\prime 2} = x^{\prime 2} + y^{\prime 2}$ and
\begin{equation}
\begin{split}
x^{\prime} & = z_{\mathrm{WCR}} \sin(i_{\mathrm{wind}}) \\ & \quad +(x_{\mathrm{WCR}} \cos(\omega_{\mathrm{wind}}) + y_{\mathrm{WCR}} \sin(\omega_{\mathrm{wind}})) \cos(i_{\mathrm{wind}}) \\
y^{\prime} & = y_{\mathrm{WCR}} \cos(\omega_{\mathrm{wind}}) - x_{\mathrm{WCR}} \sin(\omega_{\mathrm{wind}}) \\ 
z^{\prime} & = z_{\mathrm{WCR}} \cos(i_{\mathrm{wind}})\\
& \quad - (x_{\mathrm{WCR}} \cos(\omega_{\mathrm{wind}}) + y_{\mathrm{WCR}} \sin(\omega_{\mathrm{wind}}))\sin(i_{\mathrm{\mathrm{wind}}}).\\
\end{split}
\end{equation}

We can now calculate the integral in Equation\,\ref{eqn:a1}. We start from
\begin{equation*}
    a^{-1} \int n_e^2 \mathrm{d}l = \int n_e^2 \mathrm{d}z.
\end{equation*}
where we changed the integration variable to $z$, which is the distance to the source in units of the semi-major axis.

We set the lower limit of this integral to the z-coordinate of the WCR, defined as in Equation\,\ref{eq:coords}. The upper limit is set to infinity. We can now calculate the value of the optical depth using Equation\,\ref{eq:ffa} and integrating numerically, as this integral does not have an obvious analytical solution.

\begin{landscape}
\section{Flux densities used in temporal modelling}

In Table\,\ref{tab:fluxes}, we list the flux density of Apep for all frequencies and times used in our temporal modelling of the system. 

\begin{table}

\def\arraystretch{1.5}
\resizebox{\columnwidth}{!}{
    \centering
    \begin{tabular}{l|lllllllllllllllllll}
    \hline
     & 843 MHz & 1.1 GHz & 1.2 GHz & 1.4 GHz & 1.5 GHz & 1.6 GHz & 1.7 GHz & 1.8 GHz & 1.9 GHz & 2.0 GHz & 2.2 GHz & 2.3 GHz & 2.4 GHz & 2.5 GHz & 2.6 GHz & 2.7 GHz & 2.8 GHz & 3.0 GHz & 3.1 GHz \cr
    
    \hline
1988 May 05 & 85 $\pm$ 8 \cr
1989 May 09 & 71 $\pm$ 8 \cr
1990 Apr 26 & 89 $\pm$ 7 \cr
1994 Jan 10 & &&&116 $\pm$ 3 \cr
1994 Jan 22 & &&&111 $\pm$ 14 \cr
1999 May 04 & 120 $\pm$ 4 \cr
2004 Mar 21 & &&& 132 $\pm$ 5 &&&&&&&&112 $\pm$ 7  \cr
2004 Mar 23 & &&& 154 $\pm$ 4&&&&&&&&131 $\pm$ 3  \cr
2004 May 05 &  &&& 136 $\pm$ 3 &&&&&&&& 120 $\pm$ 2 \cr
2005 May 18 & 133 $\pm$ 3 \cr
2006 May 16 & 139 $\pm$ 5 \cr
2011 Jan 27 & &&& 169 $\pm$ 14 & 164 $\pm$ 5 & 145 $\pm$ 4 & 147 $\pm$ 4 & 140 $\pm$ 3 & 138 $\pm$ 3 & 134 $\pm$ 3 & 128 $\pm$ 3 & 125 $\pm$ 3 & 128 $\pm$ 3 & 124 $\pm$ 3 & 121 $\pm$ 4 & 122 $\pm$ 4 \cr
2012 Jul 03 && 167 $\pm$ 11 & 150 $\pm$ 7 & 166 $\pm$ 8 & 163 $\pm$ 6 & 149 $\pm$ 7 & 144 $\pm$ 5 & 141 $\pm$ 4 & 141 $\pm$ 4 & 140 $\pm$ 4 & 130 $\pm$ 3 & 135 $\pm$ 3 & 126 $\pm$ 3 & 125 $\pm$ 3 & 122 $\pm$ 3 & 120 $\pm$ 3 & 121 $\pm$ 3 & 116 $\pm$ 3 & 113 $\pm$ 3 \cr
2013 Jan 27 && 155 $\pm$ 13 & 154 $\pm$ 11 & 159 $\pm$ 14 & 160 $\pm$ 7 & 147 $\pm$ 8 & 151 $\pm$ 6 & 141 $\pm$ 6 & 138 $\pm$ 5 & 136 $\pm$ 4 & 135 $\pm$ 5 & 120 $\pm$ 4 & 122 $\pm$ 4 & 124 $\pm$ 4 & 121 $\pm$ 4 & 122 $\pm$ 4 & 112 $\pm$ 4 & 104 $\pm$ 4 & 105 $\pm$ 4\cr
2017 May 11 && 232 $\pm$ 20 & 199 $\pm$ 20 & 181 $\pm$ 5 & 173 $\pm$ 5 & 161 $\pm$ 6 & 158 $\pm$ 4 & 149 $\pm$ 2 & 147 $\pm$ 2 & 141.7 $\pm$ 1 & 137.9 $\pm$ 1 & 135 $\pm$ 1 & 132 $\pm$ 1 & 129 $\pm$ 2 & 126 $\pm$ 1 & 122 $\pm$ 3 & 121 $\pm$ 3 & 119 $\pm$ 3 & 114 $\pm$ 1 \cr
2018 Oct 08 && 194 $\pm$ 8 & 168 $\pm$ 7 & 170 $\pm$ 3 & 163 $\pm$ 2 & 155 $\pm$ 2 & 154 $\pm$ 2 & 151 $\pm$ 2 & 144 $\pm$ 2 & 142 $\pm$ 2 & 139 $\pm$ 2 & 133 $\pm$ 2 & 131 $\pm$ 2 & 128 $\pm$ 2 & 123 $\pm$ 3 & 122 $\pm$ 2 & 119 $\pm$ 2 & 116 $\pm$ 2 & 115 $\pm$ 2 \cr
2018 Nov 25 && 201 $\pm$ 29 & 185 $\pm$ 18 & 164 $\pm$ 2 & 163 $\pm$ 2 & 153 $\pm$ 3 & 153 $\pm$ 2 & 149 $\pm$ 2 & 143 $\pm$ 2 & 144 $\pm$ 2 & 141 $\pm$ 2 & 133 $\pm$ 2 & 133 $\pm$ 2 & 130 $\pm$ 2 & 125 $\pm$ 3 & 125 $\pm$ 3 & 121 $\pm$ 2 & 119 $\pm$ 2 & 116 $\pm$ 6 \cr
2019 Mar 24 && 181 $\pm$ 15 & 168 $\pm$ 13 & 163 $\pm$ 2 & 157 $\pm$ 3 & 149 $\pm$ 4 & 150 $\pm$ 2 & 146 $\pm$ 2 & 141 $\pm$ 2 & 139 $\pm$ 2 & 136 $\pm$ 2 & 131 $\pm$ 2 & 129 $\pm$ 2 & 126 $\pm$ 2 & 122 $\pm$ 4 & 121 $\pm$ 3 & 118 $\pm$ 2 & 114 $\pm$ 2 & 112 $\pm$ 4 \cr
2019 Sep 11 && 171 $\pm$ 14 & 161 $\pm$ 13 & 157 $\pm$ 3 & 153 $\pm$ 3 & 146 $\pm$ 5 & 147 $\pm$ 3 & 143 $\pm$ 3 & 138 $\pm$ 3 & 136 $\pm$ 3 & 132 $\pm$ 3 & 130 $\pm$ 3 & 128 $\pm$ 3 & 124 $\pm$ 3 & 122 $\pm$ 4 & 119 $\pm$ 4 & 117 $\pm$ 3 & 115 $\pm$ 3 & 112 $\pm$ 4 \cr
2020 May 02 && 183 $\pm$ 11 & 172 $\pm$ 14 & 159 $\pm$ 3 & 157 $\pm$ 3 & 149 $\pm$ 5 & 148 $\pm$ 3 & 144 $\pm$ 3 & 140 $\pm$ 3 & 137 $\pm$ 3 & 133 $\pm$ 3 & 130 $\pm$ 3 & 127 $\pm$ 3 & 126 $\pm$ 3 & 122 $\pm$ 4 & 119 $\pm$ 4 & 116 $\pm$ 4 & 114 $\pm$ 4 & 112 $\pm$ 5 \cr
2021 Mar 07 && 156 $\pm$ 14 & 145 $\pm$ 13 & 141 $\pm$ 5 & 140 $\pm$ 5 & 134 $\pm$ 6 & 132 $\pm$ 4 & 130 $\pm$ 6 & 129 $\pm$ 4 & 127 $\pm$ 4 & 125 $\pm$ 4 & 120 $\pm$ 4 & 118 $\pm$ 4 & 116 $\pm$ 4 & 112 $\pm$ 5 & 112 $\pm$ 4 & 110 $\pm$ 4 & 108 $\pm$ 4 & 103 $\pm$ 6 \cr
2021 Jun 28 && 178 $\pm$ 8 & 169 $\pm$ 12 & 161 $\pm$ 2 & 159 $\pm$ 2 & 153 $\pm$ 4 & 150 $\pm$ 2 & 148 $\pm$ 3 & 143 $\pm$ 2 & 140 $\pm$ 2 & 137 $\pm$ 3 & 134 $\pm$ 3 & 131 $\pm$ 4 & 128 $\pm$ 2 & 125 $\pm$ 4 & 122 $\pm$ 4 & 119 $\pm$ 3 & 116 $\pm$ 3 & 114 $\pm$ 4 \cr
\hline
    \end{tabular}}
    \caption{The measurements of the flux densities used in the temporal modelling in mJy. The measurements at 843\,MHz were taken with the MOST, the observations at the other frequencies were taken with the ATCA.}
    \label{tab:fluxes}
\end{table}

% Don't change these lines
\bsp	% typesetting comment
\label{lastpage}
\end{landscape}
\end{document}